\documentclass[conference]{IEEEtran}

\usepackage[utf8]{inputenc}
\usepackage[T1]{fontenc}
\usepackage{mathptmx}  
\usepackage{microtype}
\usepackage{graphicx}
\usepackage{float}
\usepackage{booktabs}
\usepackage{multirow}
\usepackage{amsmath,amssymb}
\usepackage{textcomp}
\usepackage{tabularx}
\usepackage[caption=false,font=footnotesize]{subfig}
\usepackage{algorithm}
\usepackage{algpseudocode}
\usepackage{balance}
\usepackage{hyperref}

\hypersetup{
  colorlinks=true,
  linkcolor=black,
  citecolor=black,
  urlcolor=black
}

\newcommand{\yolo}{YOLOv8}
\newcommand{\vlm}{Qwen-7B-VL}
\newcommand{\datasetN}{15,172}
\newcommand{\classesN}{16}
\newcommand{\classificationN}{11,908}

\newcommand{\segCellsN}{3,264}
\newcommand{\speedms}{1.1}
\newcommand{\fps}{909}
\newcommand{\etal}{\textit{et~al.}}

\begin{document}

\title{CAPTCHAs in the Agentic Era: Solvers That Learn from Every Encounter}

\author{
\IEEEauthorblockN{Oguzhan Salman}
\IEEEauthorblockA{Istanbul Technical University\\
\texttt{salmano21@itu.edu.tr}}
\and
\IEEEauthorblockN{Kemal Bicakci}
\IEEEauthorblockA{Istanbul Technical University\\
\texttt{kemalbicakci@itu.edu.tr}}
}

\maketitle

\begin{abstract}
  Vision-language models (VLMs) can solve visual CAPTCHAs without task-specific training, but the agents built on them approach every challenge from scratch. For such an agent, the hundredth instance of a familiar puzzle costs as much time and compute as the first. Specialized detectors invert the trade-off, answering in milliseconds but only for categories they were trained on. Neither improves with exposure. \par\smallskip We study what changes when a solver improves with use. Our system pairs a fine-tuned YOLOv8 detector with an open-weight VLM behind a confidence-based router, and runs entirely from screenshots and operating-system input events, with no browser automation or DOM access. It reaches 85.4\% overall and 84.2\% macro accuracy across 16 classes, exceeding either component alone. Every answer VLM produces also serves as a training label, so the detector absorbs categories it was never trained for, typically after one or two encounters and without human annotation. The same loop also repairs it. A CAPTCHA operator can perturb images against the publicly released detector and drive its accuracy to 0\%, but the perturbations leave VLM untouched, and its labels let the detector recover. Under a year-long simulated arms race in which the CAPTCHA operator re-crafts its perturbations each month, the solver recovers every round, and a cheap $\sim$70\%-accurate open-weight teacher hardens it as effectively as a perfect oracle. Visual CAPTCHA defenses that assume a failing bot stays failing therefore understate how quickly an adaptive solver returns.
\end{abstract}

\section{Introduction}

CAPTCHAs (Completely Automated Public Turing tests to tell Computers and
Humans Apart)~\cite{vonahn2003captcha} remain a primary defense against
automated bot activity on the web. Among modern image-based systems,
Google's reCAPTCHA v2~\cite{google2024recaptcha} is still widely deployed
despite the introduction of the implicit behavioral v3
variant of reCAPTCHA~\cite{google2024recaptchav3}. Sites continue to prefer v2 due to its explicit user verification and ease of integration into their products where visible challenge–response
mechanisms are desirable. By presenting users with semantic image labeling tasks, these challenges are intended to exploit the gap between human visual perception and current machine-vision capabilities to distinguish legitimate users from automation. However, rapid advancements in computer vision are closing this gap, challenging the security assumptions of CAPTCHAs.

Early automated solvers relied on traditional optical character recognition techniques for text-based
CAPTCHAs~\cite{bursztein2014end, yan2008lowcost}. As the CAPTCHA ecosystem
shifted toward image-based from text-based challenges, deep learning approaches emerged,
initially using convolutional neural networks trained on specific object categories and later YOLO-family detectors~\cite{redmon2016yolo}
that can achieve high accuracy on trained classes much faster than earlier approaches. Specialized YOLO models have been shown to solve all visual reCAPTCHA types --- classification (3$\times$3 grids), segmentation (4$\times$4 grids), and dynamic puzzles --- effectively~\cite{hossen2021object, plesner2024breaking} but these systems use browser automation frameworks such as Selenium, Playwright, Puppeteer, or Cypress to access the Document Object Model (DOM) to locate and interact with UI elements~\cite{selenium, playwright, puppeteer, cypress}.

Concurrently, vision-language models (VLMs) have emerged as powerful 
general-purpose systems capable of understanding and reasoning over 
diverse visual and language tasks. Building on the transformer architecture~\cite{vaswani2017attention} that revolutionized sequence modeling through self-attention mechanisms, large language models~\cite{radford2019language} demonstrated zero-shot task completion. Today, multimodal models like GPT-4V, Claude 3, and Qwen-VL drive GUI agents such as OpenAI Operator, ChatGPT Atlas, Anthropic Computer Use, WebVoyager, and Perplexity Comet~\cite{openai2025operator,openai2025atlas,anthropic2024computeruse,he2024webvoyager,perplexity2025comet} and can interact with computers in diverse tasks through user commands. Unlike specialized YOLO models, these VLM systems are capable of understanding and navigating interfaces, interpret screenshots and execute complex workflows without task specific training, which makes them primary candidate for automation. Despite this potential, recent benchmarks identify CAPTCHAs as a primary obstacle that frequently blocks these agents from completing end-to-end workflows~\cite{luo2025opencaptchaworldcomprehensivewebbased}. These challenges persist mainly because VLMs face fundamental difficulties with visual grounding and spatial localization without careful prompt engineering~\cite{yang2023setofmark}. Just as prompting strategies significantly influence
model performance on reasoning tasks~\cite{wei2022chain}, visual interactions similarly require grounding strategies. Without grounding techniques that supply localized crops and structured annotations, LLMs still struggle to identify and interact with small UI elements~\cite{omniparser}. Equipping VLMs with explicit grounding methods such as Set-of-Mark prompting bridges much of this gap, enabling strong performance in visual interaction tasks, including CAPTCHA solving~\cite{teoh2025halligan}.

From the current landscape of automated solvers, we identify three critical observations that motivate a paradigm shift in modern CAPTCHA solving. First, specialized classifiers rely heavily on browser automation frameworks (e.g., Selenium) for element localization. This both prevents integration with fully visual GUI agents and exposes the solver to detection because such frameworks leave fingerprints in the browser that anti-bot systems routinely check for. Second, while VLMs demonstrate superior reasoning capabilities and can decompose visual puzzles into step-by-step solutions~\cite{deng2024oedipus}, their computational overhead prevents deployment as real-time solvers, with inference costs orders of magnitude higher than discriminative alternatives. Third, CAPTCHA challenges represent deterministic recognition problems once the underlying visual pattern is understood, creating an opportunity to distill expensive semantic reasoning into lightweight, specialized detectors through supervised learning. This opportunity extends beyond any single CAPTCHA system---Halligan~\cite{teoh2025halligan} demonstrates that VLMs can already solve 26 distinct CAPTCHA families without task-specific training, suggesting that a distillation paradigm proven on one family could generalize broadly.

In this work, we present a hybrid solver that improves itself with use, mirroring human dual-process cognition. YOLO serves as the fast, reflexive pathway for familiar patterns, while an open-weight Qwen-7B-VL model~\cite{bai2023qwenvl} fine-tuned for web interactions~\cite{holo2025} provides the slow, deliberate reasoning for novel or ambiguous challenges. A confidence-based cascade resolves 70\% of challenges at YOLO's reflex speed (4\,ms per cell), invoking VLM reasoning (225\,ms) only when YOLO's confidence is insufficient, reducing average latency by 3 times compared to VLM-only inference. A separately fine-tuned YOLOv8 detector localizes the CAPTCHA area, grid cells, and buttons from the screenshot, with a finite-state controller coordinating detection, solving, verification, and recovery.

We make the following contributions.
\begin{itemize}\itemsep2pt \parskip0pt \topsep2pt
\item \textbf{A DOM-free hybrid solver.} The system reads screenshots and acts through OS-level mouse and keyboard events, so it neither depends on browser automation nor leaves the fingerprints such frameworks expose.

\item \textbf{A reasoning-to-reflex cascade.} YOLO answers familiar cells at reflex speed and VLM is invoked only when YOLO is uncertain, reaching 85.4\% overall and 84.2\% macro accuracy, higher than either model alone.

\item \textbf{Autonomous self-extension and recovery.} Because every challenge VLM solves becomes training data, the solver absorbs object classes it was never shipped with, often from one or two puzzle encounters, and recovers on its own when a PGD attack collapses YOLO from 83\% to 0\%, in both cases without human annotation.

\item \textbf{Adaptive-robustness findings.} Under a year-long arms race in which the CAPTCHA operator re-crafts its perturbations each month, the solver re-hardens every round, and a cheap $\sim$70\%-accurate teacher proves as effective for this hardening as a perfect oracle, since its labeling errors make the solver harder to replicate.
\end{itemize}
\section{Related Work}
\begin{table*}[t]
\centering
\caption{Modern CAPTCHA solvers: macro accuracy, robustness under attack, and operational cost.}
\label{tab:comparative_analysis_between_papers}
\footnotesize
\begin{tabularx}{\textwidth}{@{}l l c c X l@{}}
\toprule
\textbf{Paper} & \textbf{Primary Model} & \textbf{DOM-Free} & \textbf{Macro Acc.} & \textbf{Robustness} & \textbf{Cost} \\
\midrule
Weng \etal~\cite{weng2019towards} & Specialized CNNs & \texttimes & N/A & Not evaluated & On-premises \\ \addlinespace
Hossen \etal~\cite{hossen2021object} & YOLOv3 (Detection) & \texttimes & 60.74\%$^*$ & 73\% on reCAPTCHA-perturbed images (corruption only; no PGD) & On-premises \\ \addlinespace
Plesner \etal~\cite{plesner2024breaking} & YOLOv8 (Classification) & \texttimes & 75.69\%$^*$ & 0\% under PGD ($\epsilon$=4/255); no adversarial training evaluated & On-premises \\ \addlinespace
OEDIPUS~\cite{deng2024oedipus} & GPT-4 / Gemini & \texttimes & N/A & Not evaluated & \$3--13 / 100 \\ \addlinespace
Halligan~\cite{teoh2025halligan} & GPT-4o & Partial & N/A & $-$15--49\,pp under visual distractors (no PGD) & \$2.40 / 100 \\ \addlinespace
\midrule
\textbf{Our System} & \textbf{Qwen-7B-VL + YOLOv8} & \checkmark & \textbf{84.17\%} & \textbf{VLM: 75\% under PGD ($\epsilon$=4/255); YOLO recovers to 65\% via VLM-labeled retraining (GT: 71\%)} & \textbf{On-premises} \\
\bottomrule
\end{tabularx}
\vspace{1mm} \\
\scriptsize $^*$ Calculated from per-class success rates provided in the original paper's figures. N/A = not available.
\end{table*}
Early automated CAPTCHA solvers focused on text-based challenges using OCR and segmentation techniques~\cite{mori2003recognizing, yan2008lowcost, bursztein2014end}, which ultimately led to the development of image-based systems like reCAPTCHA v2. Sivakorn \etal~\cite{sivakorn2016breaking} demonstrated large-scale attack on reCAPTCHA v2, achieving 70.78\% success rate on image challenges using deep learning-based image annotation services and 83.5\% on Facebook's image CAPTCHA. Subsequent work targeted specifically image-based reCAPTCHA systems using deep convolutional networks trained on specific object categories. Hossen \etal~\cite{hossen2021object} demonstrated that YOLOv3-based object detection can solve Google's image reCAPTCHA v2 with 83.25\% success rate, averaging 19.93 seconds per CAPTCHA. Their approach used a bounding box to grid mapping algorithm to convert YOLOv3 detections into cell selections for 3$\times$3 classification puzzles. While effective for COCO-trained object classes (bicycles, buses, cars, traffic lights), their system required puppeteer-firefox for DOM access to extract grid cell coordinates and trigger click events. Hossen \etal~also evaluated robustness against reCAPTCHA's built-in image perturbations (noisy, distorted challenge images), achieving 73\% object detection on corrupted samples after adversarial training, though they did not consider gradient-based attacks. Weng \etal~\cite{weng2019towards} demonstrated systematic attacks across three image CAPTCHA types (selection-based, slide-based, click-based) using CNNs and Fast-RCNN, achieving 79--90\% success on selection-based puzzles similar to reCAPTCHA. Plesner \etal~\cite{plesner2024breaking} fine-tuned YOLOv8 on 14,000 reCAPTCHA images across 13 object classes, achieving near-perfect solving rates on image classification puzzles. However, Plesner \etal~ system also relies on Selenium for browser automation and remains limited to trained object classes, achieving 0\% coverage on unseen categories. 

These prior CAPTCHA solvers share a critical limitation: reliance on browser automation frameworks (Selenium, Playwright) to access the Document Object Model (DOM) for element localization and interaction. This dependency matters for two reasons. First, it is incompatible with the trend toward fully visual GUI agents that interact with computers through screenshots and OS-level input events without DOM access~\cite{anthropic2024computeruse}. A CAPTCHA solver that requires DOM access cannot serve as a module within such agents. Second, browser automation frameworks inject detectable artifacts into the browser environment --- Selenium, for example, sets the \texttt{navigator.webdriver} property --- and anti-bot systems routinely check for such fingerprints~\cite{garcia2024browser}. Plesner~\etal~\cite{plesner2024breaking} compensate for this detectability through VPN rotation, browser cookies, and B\'{e}zier-curve mouse trajectories; without these, their solver degrades significantly (Section~\ref{sec:live_eval}). Our system avoids this class of detection by operating at the OS level, capturing screenshots via system APIs and clicking at screen coordinates via PyAutoGUI, with no browser API or DOM access at any point.

To overcome these limitations, research has shifted toward purely visual interface understanding, treating the browser as a visual canvas rather than a structured document. In this broader context, YOLO-family detectors have been extensively used for UI element detection in mobile and web interfaces, achieving reliable localization of buttons, icons, and other general components. OmniParser~\cite{omniparser} further demonstrated that pretrained YOLO detectors can accurately localize even small UI icons, but it did not consider CAPTCHA layouts or puzzle-specific grid structures. In parallel, multimodal vision--language models such as CLIP~\cite{radford2021clip}, Flamingo~\cite{alayrac2022flamingo}, LLaVA~\cite{liu2023llava}, and InstructBLIP~\cite{dai2023instructblip} have greatly improved high-level screen understanding and visual reasoning. Yet prior evaluations of VLM-based GUI agents~\cite{luo2025opencaptchaworldcomprehensivewebbased} (e.g., GPT-4V, Qwen-VL, Claude 3) report poor performance on CAPTCHAs, reflecting limited fine-grained grounding and a lack of structured output mechanisms needed for grid-style selection tasks. Together, these results suggest that neither generic UI detectors nor current VLM agents alone directly address CAPTCHA-specific perception and solution.

Beyond browser-based automation frameworks, Zhang \etal~\cite{zhang2024ufo} proposed UFO, a Windows UI-automation agent that interacts with applications via the Windows UI Automation API instead of Selenium. Although it removes the dependency on a web browser, UFO still assumes programmatic accessibility hooks and thus does not operate in a purely screenshot-based setting.

While these agents focus on the mechanics of interaction, other recent works address the logical reasoning required to solve complex puzzles. Deng \etal~\cite{deng2024oedipus} successfully employs a Domain Specific Language (DSL) to decompose complex reasoning CAPTCHAs into substeps and achieves an average success rate of 63.5\% across multiple reasoning categories; however, it faces significant latency (>100 seconds) and cost constraints for every execution. Our work shifts this paradigm and, rather than using VLM as a standalone inference solver for every challenge, we propose a hybrid teacher-student model, addressing the fundamental economic and speed bottlenecks identified in the OEDIPUS framework.

Most recently, Teoh \etal~\cite{teoh2025halligan} introduced Halligan, a VLM-based CAPTCHA solver (GPT-4o~\cite{openai2023gpt4v}) that achieves a 60.7\% success rate across 26 CAPTCHA families and 68.0\% on reCAPTCHA v2 specifically. Halligan formulates CAPTCHA solving as a search problem where a vision-language model iteratively explores candidate actions guided by natural language objectives. While Halligan operates in a DOM-free environment, its reliance on iterative search results in a median latency of 21.8 seconds per challenge and an operational cost of \$2.40 per 100 solutions. Furthermore, Halligan reports that solve rates drop by 15.4\% to 49.6\% when faced with adversarial transformations or distractions such as gaussian noise and advertisement banners and assumes the challenge can be cleanly isolated from the surrounding interface. 
In contrast, our approach eliminates the assumption of clean isolation by performing end-to-end learned visual grounding via a fine-tuned YOLOv8 backbone, enabling robust handling of cluttered layouts and dynamic variants. 

Table~\ref{tab:comparative_analysis_between_papers} compares accuracy, robustness, and cost across modern reCAPTCHA v2 solvers and reasoning agents. Specialized classifiers~\cite{hossen2021object, plesner2024breaking} are accurate on the classes they were trained for but have no coverage beyond them, and they rely on browser automation frameworks vulnerable to DOM-based detection, requiring substantial behavioural infrastructure to operate at scale. Robustness is largely unaddressed across prior work, with Plesner's classifier dropping to 0\% under PGD and Halligan losing 15--49\,pp under visual distractors alone. Generalized agentic frameworks~\cite{teoh2025halligan, deng2024oedipus} handle broader task distributions but at order-of-magnitude higher per-puzzle cost. Our system pairs Plesner's published classifier with an open-weight VLM cascade, raising macro accuracy across the full 16-class distribution to 84.17\%, retaining 75\% under PGD through VLM path, and avoiding DOM-based detection without per-call API costs. Because both systems use the same published YOLO classifier, these gains come from the hybrid design rather than a stronger detector.

\section{Method}

\begin{figure*}[t]
  \centering
  \includegraphics[width=\textwidth]{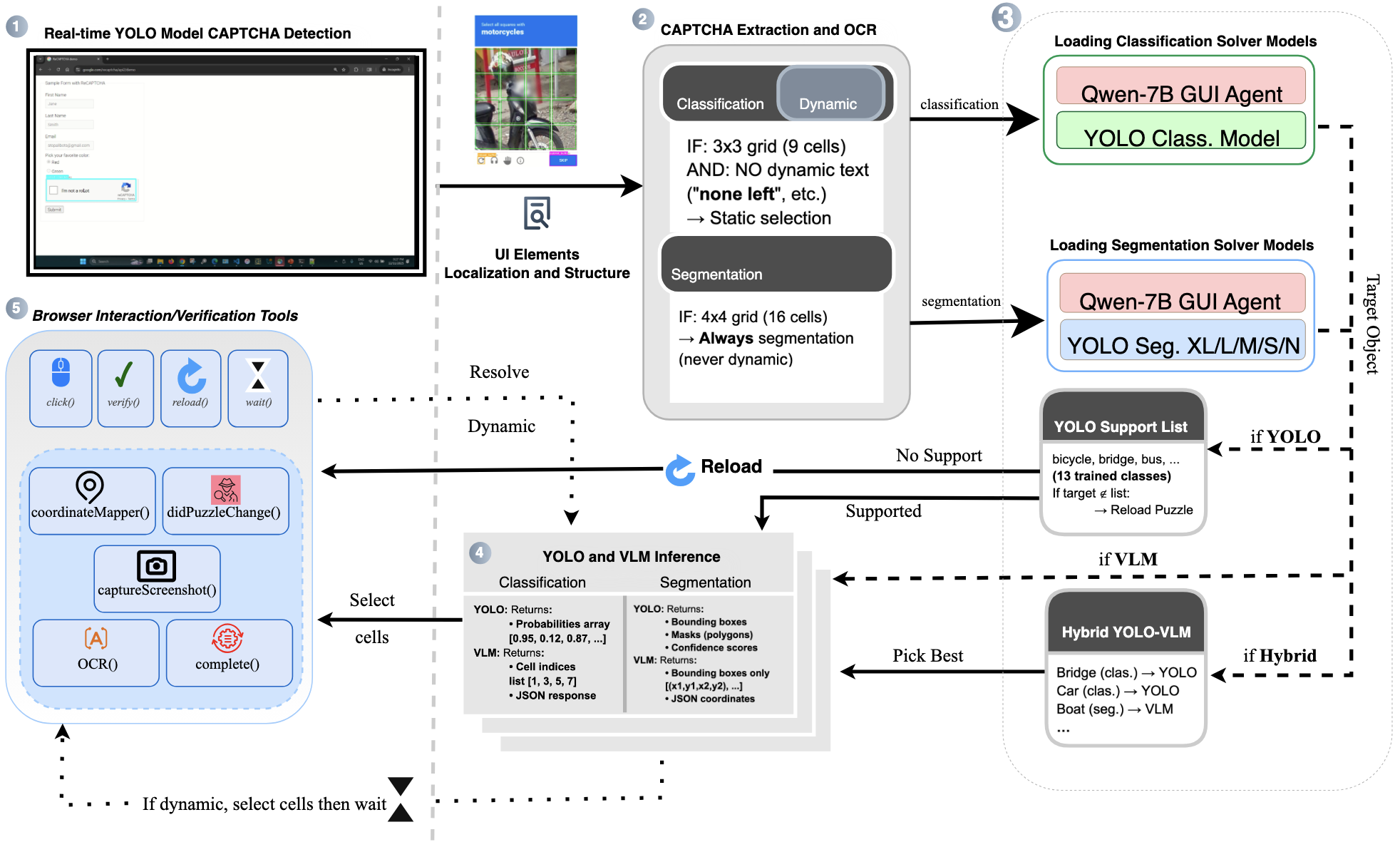}
  \caption{Complete system architecture showing end-to-end CAPTCHA solving pipeline with execution steps. \textbf{Step 1:} Our system captures screenshots and uses a fine-tuned YOLOv8 detector to localize UI elements (captcha area, grid cells, verify/reload buttons). \textbf{Step 2:} Once UI elements are detected, OCR extracts the target object (e.g., "select all motorcycles") and determines the puzzle type based on the detected number of grid cells (classification vs.\ segmentation). \textbf{Step 3:} Based on puzzle type, the system loads appropriate solver models. The hybrid routing strategy assigns each class to YOLO or VLM based on the performance for the target class. If solver engine is YOLO and the target is unsupported by YOLO (e.g., Boat, Taxi, Tractor), the system reloads the puzzle. \textbf{Step 4:} YOLO returns probability arrays (classification) or masks/bounding boxes (segmentation); VLM returns cell indices or bounding box coordinates in JSON format. \textbf{Step 5:} The finite-state machine maps detected UI elements from screenshot pixel coordinates to absolute screen coordinates and sends OS-level mouse and keyboard events; the browser is never accessed programmatically. It then verifies success via screenshot capture, reloads the puzzle on failure, and handles dynamic puzzles where cells refresh after interaction.}
  \label{fig:system_architecture}
\end{figure*}

In this section, we present our hybrid architecture designed to solve visual CAPTCHA challenges without DOM access. The proposed system consists of a fine-tuned object detector for UI localization, a deterministic finite-state machine (FSM) for process control and a dual-path solving mechanism that leverages the complementary strengths of specialized YOLO classifiers and Vision-Language Models. The entire pipeline operates at the OS level: input is obtained by capturing screenshots via system APIs, and output is delivered by sending mouse and keyboard events to absolute screen coordinates. The browser is treated as a visual surface and is never accessed programmatically. Figure~\ref{fig:system_architecture} illustrates the complete system architecture with execution steps.

Our architecture builds on two complementary model families. You Only Look Once (YOLO)~\cite{redmon2016yolo} is a family of object detectors that, given an input image $x \in \mathbb{R}^{H \times W \times 3}$, produces bounding box predictions $\{(b_i, c_i, p_i)\}_{i=1}^{K}$ in a single forward pass, where $b_i = (x, y, w, h)$ defines coordinates, $c_i \in \{1, \dots, C\}$ is the predicted class, and $p_i \in [0,1]$ is the confidence score. For classification, YOLO produces a probability distribution over $C$ classes:
\begin{equation}
    f(x) = \text{softmax}(z), \quad z_j = w_j^\top \phi(x) + b_j
\end{equation}
where $\phi(x)$ denotes the feature representation and $w_j, b_j$ are the weights and biases for class $j$. In our system, we employ three separately trained YOLO models for UI element detection, tile classification, and instance segmentation respectively.

Vision-Language Models (VLMs) complement YOLO's discriminative classification with open-ended reasoning. VLM maps an image $x$ to visual tokens $v = g_v(x) \in \mathbb{R}^{N_v \times d}$ via a vision encoder, then generates output tokens autoregressively:
\begin{equation}
    P(y_1, \dots, y_T \mid v, t) = \prod_{i=1}^{T} P_\theta(y_i \mid y_{<i}, v, t)
\end{equation}
This token-by-token generation enables flexible reasoning but costs substantially higher latency than YOLO's single-pass inference. We use Qwen-7B-VL~\cite{bai2023qwenvl}, a 7-billion parameter VLM fine-tuned for GUI interaction tasks~\cite{holo2025}, constraining its output to a single class index for classification or bounding box coordinates for segmentation.

\subsection{Detection Backbone: \yolo}

We fine-tune YOLOv8 from OmniParser's~\cite{omniparser} pretrained icon detection model~\cite{omniparsergithub} to detect five CAPTCHA-specific classes: captcha area, cell, robot checkbox, reload button, and submit button. OmniParser provides strong foundation model for small UI element localization, trained on diverse webpage screenshots. To create a diverse training dataset without exposing our model to live CAPTCHA services during development, we synthetically generated 1,200 training samples by programmatically overlaying CAPTCHA elements onto webpage backgrounds from a Kaggle screenshot dataset~\cite{kaggle_screenshots} (details in Appendix B.2). The model processes 512$\times$512 RGB images and we train with AdamW optimizer, multi-scale training and early stopping following established fine-tuning practices.

\subsection{Finite-State Controller}

We coordinate detection, solving, and verification through a deterministic finite-state machine $\mathcal{M} = (S, s_0, \Sigma, \delta)$ with states $S$ covering initialization, puzzle analysis, solving, verification, and recovery. Transitions $\delta : S \times \Sigma \rightarrow S$ are triggered by visual events (checkbox detected, grid appeared, CAPTCHA cleared). The FSM handles both static and dynamic puzzles. In static puzzles, the system solves once and then verifies. In dynamic puzzles, selected tiles refresh with new images after each click, requiring the FSM to iteratively re-analyze the grid until no target objects remain and then clicks verify.
 The complete solving procedure integrating FSM control with hybrid YOLO/VLM solving is presented in Algorithm~\ref{alg:pixelcap} in the Appendix.
\subsection{Classification and Segmentation Pipelines}
\label{subsec:solving_paths}

After extracting all CAPTCHA-related UI elements using our trained YOLOv8 detector, we use EasyOCR to detect the instruction text within the CAPTCHA area and extract the target class. With the target identified, individual cells cropped from the detected captcha area and images are sent either to a YOLO classifier/segmentation~\cite{plesnergithub} or to Qwen-7B-VL GUI agent~\cite{hologithub}.

For classification puzzles, VLM receives a numbered prompt listing all 16 classes (0--15) and is constrained to return a single class index, enforcing deterministic top-1 predictions and minimizing hallucinations (exact prompt in Appendix~\ref{sec:prompts}). Each of the 9 cells is extracted from the 3$\times$3 grid and independently classified by either YOLO or VLM (see Figure~\ref{fig:classification_pipeline} in the Appendix).

For segmentation puzzles, 4$\times$4 detected cells are concatenated into a single image. YOLO produces masks or bounding boxes, while VLM receives a structured prompt requesting bounding box coordinates for all instances of the target object (exact prompt in Appendix~\ref{sec:prompts}). Predictions are mapped back to individual cell coordinates using a 1\% overlap threshold: any cell whose bounding box intersects the predicted region by $\geq$1\% of its area is selected (see Figure~\ref{fig:segmentation_pipeline} in the Appendix).


\section{Data and Metrics}
To accurately measure the performance of our hybrid solver, we constructed a comprehensive dataset reflecting real-world class distributions that contain challenging edge cases. This section details the composition of our dataset and defines the specific evaluation metrics used to measure performance across both classification and segmentation puzzle types.
\subsection{Dataset Composition}
Our evaluation dataset comprises \datasetN{} samples in total, collected from public reCAPTCHA dataset \cite{mandourah2024recaptcha} and CAPTCHA samples obtained during our experiments. The dataset covers \classesN{} object classes: Bicycle, Boat, Bridge, Bus, Car, Chimney, Crosswalk, Hydrant, Motorcycle, Mountain, Other, Palm, Stairs, Taxi, Tractor, and Traffic Light. Among these classes, three of these (Boat, Taxi, Tractor) are not supported by the published YOLO classifier. We encountered these object categories during our experiments and include them in our evaluation to reflect real-world scenarios. Because the published YOLO classifier was itself trained on part of this dataset, we evaluate exclusively on held-out data that neither model has seen.

\subsection{Evaluation Metrics}
\label{subsec:evaluation_metrics}
We use different primary metrics for classification and segmentation puzzles based on their task requirements. For 3$\times$3 classification puzzles, we prioritize recall because these types of CAPTCHA puzzles require identifying all tiles that contain target object and missing a required tile increases risk score. 

We report both macro-averaged recall and overall accuracy to capture different aspects of solver robustness. Overall accuracy reflects performance on the natural class distribution in our dataset, where some classes appear more frequently than others. However, from a security perspective, macro-averaged metrics are more critical: CAPTCHA designers can adaptively bias puzzle distribution toward object classes where automated solvers are known to fail. For instance, if YOLO-based solvers cannot recognize Boats, Taxis, or Tractors, defenders could increase the frequency of these categories to maximize solver failure rate. Macro-averaged recall treats all classes equally regardless of frequency, providing a worst-case performance measure that is robust to such adversarial class selection. A solver with high overall accuracy but low macro-averaged recall remains vulnerable to targeted countermeasures, whereas our hybrid approach achieves strong performance on both metrics demonstrating resilience against adaptive defenses.

Our evaluation for the classification task differs from live sessions of CAPTCHA puzzles in that we assess models on a static dataset of puzzle tiles from all 16 classes, requiring the model to discriminate the dominant object across all possible categories. In contrast, real reCAPTCHA presents context-specific puzzles where positive examples contain obvious instances of the target and negative examples contain clearly different objects. This makes our evaluation more challenging than live solving because models must handle images that contain multiple categories such as distinguishing dominant bridges from dominant cars in images where both may be present as occluded, out of context or secondary elements.


In this static evaluation setting, recall reliably measures whether the model identifies the ground-truth dominant object, while precision can be artificially deflated when models detect genuine but secondary or occluded objects in the scene, (see Figure~\ref{fig:boats_inside_car} in the Appendix). We report per-class precision, recall, and F1-score, along with overall accuracy (frequency-weighted, where frequent classes dominate) and macro-averaged accuracy (where all classes contribute equally).

An alternative design would be to report top-$k$ accuracy or
probability-threshold metrics (e.g., treating a prediction as correct if
Car appears in the top-3 classes with probability at least
$10\%$). We deliberately refrain from doing so in order to maintain a
fair evaluation between YOLO and VLM. The YOLO
classifier outputs a full softmax distribution over classes, so
top-$k$ and probability thresholds are well defined.
VLM, in contrast, does not expose class probabilities:
in our setup, it produces a single class index token (0--15), and when
instructed to output probabilities, it returns unnormalized,
inconsistent or even hallucinated values that often do not sum to $100\%$. These values reflect the model's language-generation behaviour rather than reliable posterior probabilities and there is no principled way to construct a top-$k$ metric for VLM that is comparable to YOLO's softmax scores. To avoid this asymmetry, we discard the additional confidence information provided by YOLO and evaluate both models under the shared metric which is single top-1 label per tile.

For 4$\times$4 segmentation puzzles, we use F1-score as the primary
metric, since boundary localization requires balancing both
under-selection (low recall) and over-selection (low precision),
making F1-score appropriate for measuring segmentation performances.

\section{Experiments}
\label{sec:experiments}
We evaluate discriminative and generative models through a series of experiments designed to highlight their distinct capabilities. We analyze classification performance in detail, summarize the extension to segmentation, and conclude with a latency analysis of each model.

\subsection{Classification Experiments}
To evaluate the solvers on classification-type puzzles, we compare the published YOLO classifier and our Qwen-7B-VL GUI agent on identical reCAPTCHA challenges. This analysis focuses on cases where each grid cell is to be classified independently.

Following the evaluation protocol described in
Section~\ref{subsec:evaluation_metrics}, we evaluate both models on a
held-out test set spanning \classesN{} object
classes, requiring a single top-1 prediction per image.
Because the published YOLO classifier was trained on the training split
of the reCAPTCHA dataset~\cite{mandourah2024recaptcha}, we construct the
test set from the validation split plus separately collected samples for
four classes (Boat, Stairs, Taxi, Tractor) that were not part of YOLO's
training data. Neither model was trained on any image in this test set.
For YOLO, we take the argmax over its softmax scores and for VLM, we
use the numbered 0--15 prompt described in Section~\ref{subsec:solving_paths},
which constrains it to return exactly one class index.

\begin{figure*}[t]
  \centering
  \subfloat[YOLO normalized confusion matrix.\label{fig:yolo_cm}]{%
    \includegraphics[width=0.48\textwidth]{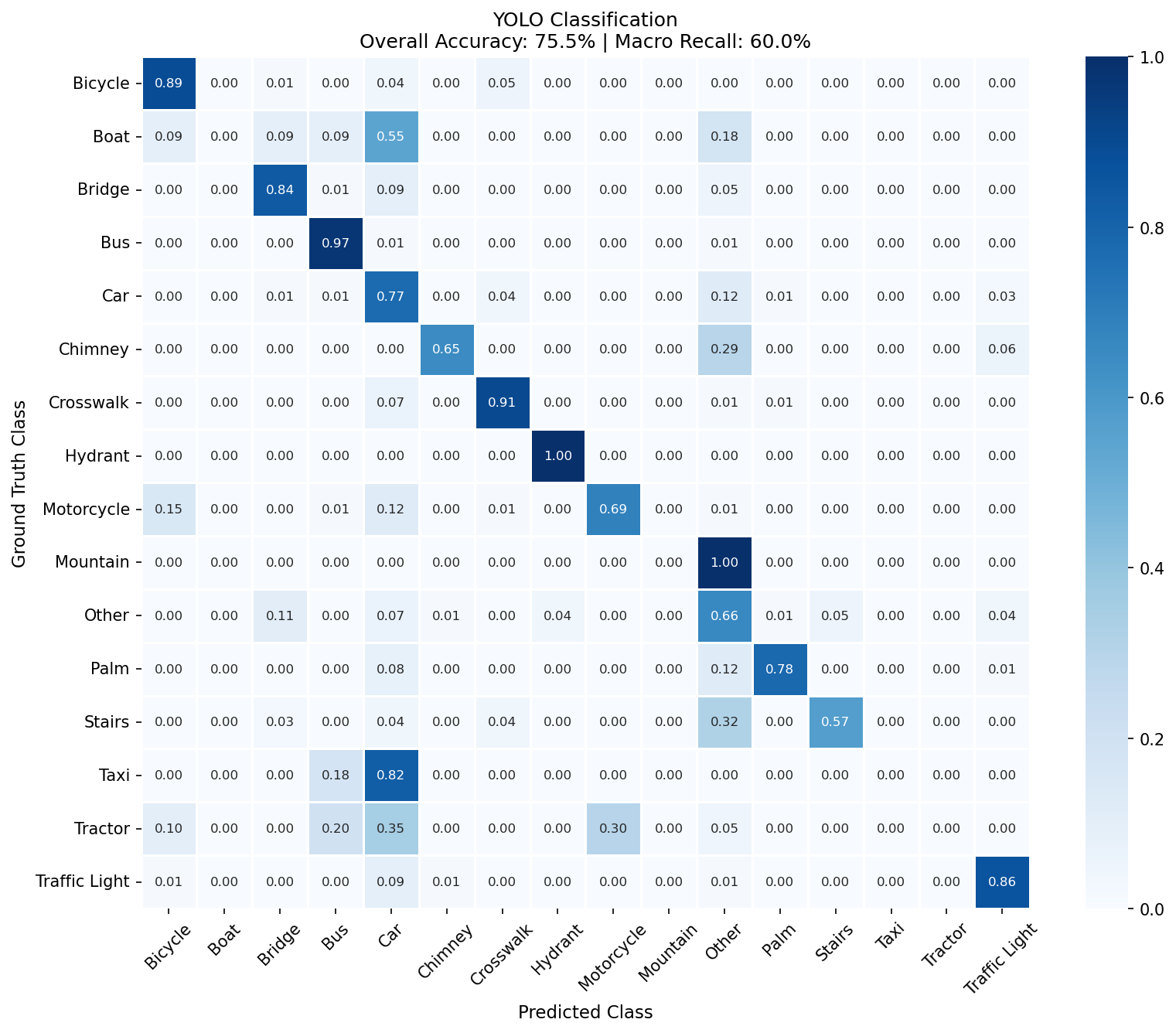}}
  \hfill
  \subfloat[GUI Agent (VLM) normalized confusion matrix.\label{fig:vlm_cm}]{%
    \includegraphics[width=0.48\textwidth]{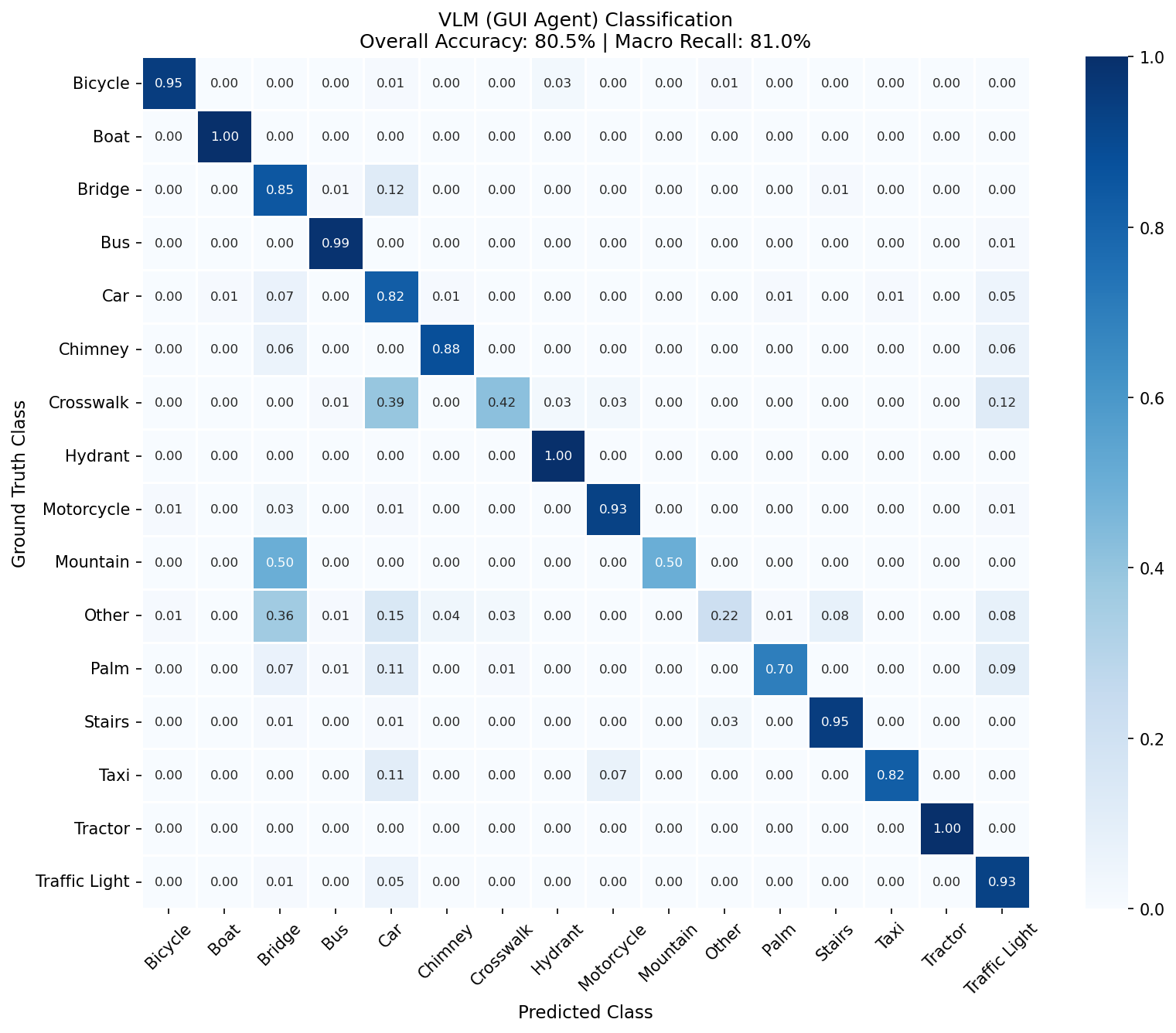}}
  \caption{Comparison of per-class recall for YOLO vs VLM GUI agent.
  Darker diagonals indicate stronger recall; off-diagonals correspond
  to confusion.}
  \label{fig:class_confusion}
\end{figure*}

The test set exhibits class imbalance reflective of real-world
reCAPTCHA distributions. Most classes contain 74 validation images, while
rare classes (e.g., Boat: 11, Mountain: 2, Chimney: 17) have fewer samples.

\paragraph{YOLO Classifier Performance} 
We first evaluate the published YOLO classifier~\cite{plesner2024breaking}. As shown in Table~\ref{tab:classification_metrics}, the model achieves an overall accuracy of 75.5\% but a macro accuracy of only 60.0\% across all 16 classes. We also report 12-class metrics excluding the four separately collected classes (Boat, Stairs, Taxi, Tractor) for comparability with prior work that evaluated only on the original dataset classes. On these 12 classes, YOLO achieves 83.1\% top-1 classification accuracy, closely matching Plesner~\etal's reported 82.4\% top-1 accuracy on the same model (the small difference likely reflects additional evaluation samples from their live bot operations that are not publicly available).
Looking at the per-class performances in Table~\ref{tab:perclass}, YOLO achieves high recall on classes such as \textit{Bus} (97.3\%), \textit{Hydrant} (100.0\%) and texture-heavy classes like \textit{Crosswalk} (90.5\%). However, it struggles with structurally similar objects like \textit{Motorcycles} (68.9\% recall), which are frequently misclassified as Bicycles (Figure~\ref{fig:yolo_cm}), and scores 0\% on unsupported classes.

\begin{table}[H]
\centering
\caption{Classification accuracy and latency comparison on reCAPTCHA object categories.}
\label{tab:classification_metrics}
\begin{tabular}{lccc}
\toprule
Metric & YOLO & VLM (GUI Agent) & Hybrid \\
\midrule
Overall Acc. (16 classes) & 0.7548 & 0.8052 & 0.8544 \\
Overall Acc. (12 classes$^*$) & 0.8314 & 0.7827 & 0.8498 \\
Macro Acc. (16 classes)   & 0.5999 & 0.8101 & 0.8417 \\
Macro Acc. (12 classes$^*$)   & 0.7521 & 0.7661 & 0.8160 \\
\midrule
VLM calls & 0\% & 100\% & 30.0\% \\
Avg. latency (per cell) & 4\,ms & 225\,ms & 71\,ms \\
\bottomrule
\multicolumn{4}{l}{\scriptsize $^*$Excludes four separately collected classes (Boat, Stairs, Taxi, Tractor).}
\end{tabular}
\end{table}

\begin{table*}[!t]
\centering
\caption{Per-class Precision, Recall, and F1 scores.}
\label{tab:perclass}
\small
\begin{tabular}{l r r r r r r}
\toprule
\multirow{2}{*}{Class} & \multicolumn{3}{c}{YOLO} & \multicolumn{3}{c}{VLM (GUI Agent)} \\
\cmidrule(lr){2-4} \cmidrule(lr){5-7}
& Prec & Rec & F1 & Prec & Rec & F1 \\
\midrule
Bicycle      & 81.5\% & 89.2\% & 85.2\% & 97.2\% & \textbf{94.6\%} & 95.9\% \\
Boat$^\dagger$ & 0.0\% & 0.0\% & 0.0\% & 91.7\% & \textbf{100.0\%} & 95.7\% \\
Bridge       & 82.7\% & 83.8\% & 83.2\% & 59.4\% & \textbf{85.1\%} & 70.0\% \\
Bus          & 84.7\% & 97.3\% & 90.6\% & 94.8\% & \textbf{98.6\%} & 96.7\% \\
Car          & 41.0\% & 77.0\% & 53.5\% & 47.7\% & \textbf{82.4\%} & 60.4\% \\
Chimney      & 84.6\% & 64.7\% & 73.3\% & 78.9\% & \textbf{88.2\%} & 83.3\% \\
Crosswalk    & 85.9\% & \textbf{90.5\%} & 88.2\% & 91.2\% & 41.9\% & 57.4\% \\
Hydrant      & 96.1\% & 100.0\% & 98.0\% & 94.9\% & 100.0\% & 97.4\% \\
Motorcycle   & 89.5\% & 68.9\% & 77.9\% & 94.5\% & \textbf{93.2\%} & 93.9\% \\
Mountain     & 0.0\% & 0.0\% & 0.0\% & 100.0\% & \textbf{50.0\%} & 66.7\% \\
Other        & 45.0\% & \textbf{66.2\%} & 53.6\% & 84.2\% & 21.6\% & 34.4\% \\
Palm         & 95.1\% & \textbf{78.4\%} & 85.9\% & 96.3\% & 70.3\% & 81.2\% \\
Stairs       & 91.5\% & 57.3\% & 70.5\% & 91.0\% & \textbf{94.7\%} & 92.8\% \\
Taxi$^\dagger$ & 0.0\% & 0.0\% & 0.0\% & 95.8\% & \textbf{82.1\%} & 88.5\% \\
Tractor$^\dagger$ & 0.0\% & 0.0\% & 0.0\% & 100.0\% & \textbf{100.0\%} & 100.0\% \\
Traffic Light & 90.1\% & 86.5\% & 88.3\% & 70.4\% & \textbf{93.2\%} & 80.2\% \\
\midrule
Macro Avg. & 60.5\% & 60.0\% & 59.3\% & 86.8\% & 81.0\% & 80.9\% \\
\bottomrule
\multicolumn{7}{l}{\scriptsize $^\dagger$Unsupported by YOLO Classification Model}
\end{tabular}
\end{table*}

\paragraph{VLM GUI Agent Performance} 

Next, we evaluate our \vlm{} GUI agent. VLM achieves an overall accuracy of 80.5\% and macro accuracy of 81.0\%.
Unlike YOLO, VLM demonstrates strong zero-shot generalization, achieving 100\% recall on rare classes like Boat and Tractor and high recall on semantically distinct categories like Traffic Light (93.2\%) and Stairs (94.7\%). However, VLM underperforms on specific categories. For example, VLM often ignores the primary object like Crosswalk to focus on secondary objects (classifying them as Bridges or Others or hallucinates specific labels for generic backgrounds).

\paragraph{Hybrid Routing Strategy}

Table~\ref{tab:classification_metrics} summarizes the performance of both models alongside our hybrid routing strategy.

Rather than statically assigning each class to a single model, we employ a confidence-based cascade that routes each image independently. For unsupported classes (identified via OCR from the puzzle instruction), images are routed directly to VLM. For supported classes, YOLO processes the image first and its top-1 softmax confidence determines whether to trust its prediction or fall back to VLM:
\begin{equation}
\small
r(x, c) \!=\! \begin{cases}
f_{\text{VLM}}(x) & c \notin \mathcal{C}_{\text{YOLO}} \\
f_{\text{YOLO}}(x) & c \in \mathcal{C}_{\text{YOLO}} \!\wedge\! \max f_{\text{YOLO}}(x) \!\geq\! \tau \\
f_{\text{VLM}}(x) & c \in \mathcal{C}_{\text{YOLO}} \!\wedge\! \max f_{\text{YOLO}}(x) \!<\! \tau
\end{cases}
\label{eq:cascade}
\end{equation}
where $c$ is the target class, $\mathcal{C}_{\text{YOLO}}$ is the set of YOLO-supported classes, and $\tau$ is the confidence threshold. 


Table~\ref{tab:classification_metrics} functions as a controlled ablation over the same test set and metrics, isolating the contribution of each component: YOLO-only, VLM-only, and the hybrid are evaluated on identical images so that the differences are attributable to the routing design alone. The two single-model configurations are each insufficient in a distinct way. YOLO-only reaches 75.5\% overall but collapses to 60.0\% macro accuracy because it scores 0\% on the three unsupported classes; VLM-only lifts macro accuracy to 81.0\% through zero-shot coverage but is slower and less precise on the supported classes that dominate the distribution. With $\tau = 0.70$, the hybrid achieves 85.4\% overall and 84.2\% macro accuracy, exceeding \emph{both} single-model baselines rather than merely interpolating between them: it improves on YOLO-only by +10.0\,pp overall and +24.2\,pp macro, and on VLM-only by +4.9\,pp overall and +3.2\,pp macro, while invoking VLM for only 30\% of images.

Importantly, this advantage is not merely an artifact of YOLO scoring 0\% on the three unsupported classes. On the 12-class subset that excludes the separately collected classes---where every class is one YOLO was trained to support---the hybrid still exceeds both single models, reaching 85.0\% overall (vs.\ YOLO's 83.1\% and VLM's 78.3\%) and 81.6\% macro (vs.\ 75.2\% and 76.6\%). Here the gain comes entirely from the confidence-based fallback: when YOLO's top-1 confidence is low on a supported class, routing that image to VLM corrects errors that neither model avoids alone. The average per-cell latency is 71.9\,ms, a 3.1$\times$ reduction compared to VLM-only inference. Table~\ref{tab:perclass} presents per-class Precision, Recall, and F1 scores.

\subsection{Segmentation}

The hybrid architecture also extends to 4$\times$4 segmentation puzzles, where every cell overlapping the target object must be selected. As in classification, a hybrid (YOLOv8x-mask + VLM) performs best, reaching 88.6\% accuracy (F1=0.850) by combining YOLO's geometric precision on supported classes with VLM's zero-shot coverage of unsupported ones. Unlike classification, however, the 7B VLM is not an adequate teacher here: its segmentation predictions (F1=0.769) are less precise than even a mid-sized YOLOv8m (F1=0.811), so distilling them would degrade rather than improve YOLO. We therefore restrict the teacher-student and adversarial experiments to classification (full per-model and per-class segmentation results in Appendix~\ref{app:segmentation}).

\subsection{Latency Analysis}
Table~\ref{tab:latency} presents the latency analysis for different models in different puzzle types. The YOLOv8 detector operates in real-time at \speedms{} ms per image, enabling immediate UI localization.
\begin{table}
\centering
\caption{Component Latency (median; ms per inference)}
\label{tab:latency}
\begin{tabular}{lcc}
\toprule
Component & Latency (ms) & FPS \\
\midrule
YOLOv8 Detector & \speedms & \fps \\
YOLO Classifier (per cell) & 4.4 & 220 \\
YOLOv8n-seg (per grid) & 23 & 43.5 \\
YOLOv8x-seg (per grid) & 26 & 38.2 \\
VLM Classification (per cell) & 225 & 4.1 \\
VLM Segmentation (per grid) & 2251 & 0.4 \\
\bottomrule
\end{tabular}
\end{table}
A significant performance gap exists between specialized YOLO solvers and VLM. The YOLO-only pipeline maintains real-time performance ($\approx$26\,ms for segmentation) because YOLO is architecturally optimized for real-time inference with lightweight size ($<$100M parameters). In contrast, VLM exhibits a substantial latency cost (225\,ms--2251\,ms) driven by two factors. First, the Qwen-7B-VL model (7 billion parameters) requires significantly higher compute and memory bandwidth per inference than the YOLO models. Second, unlike YOLO's near-instantaneous inference, VLM must generate text tokens sequentially through autoregressive generation. This cost is particularly evident in the segmentation task (2251\,ms), where VLM must generate a long sequence of coordinate tokens for every detected object, whereas classification requires generating only a single class index token (225\,ms).


\subsection{Discussion}
Our results show that these models demonstrate fundamentally different recognition strategies such that YOLO relies on learned silhouettes, while VLM leverages semantic reasoning from internet-scale pre-training. For instance, YOLO frequently confuses Motorcycles with Bicycles, with 15\% of Motorcycle images misclassified as Bicycles. This confusion likely stems from both severe training imbalance (27 Motorcycle samples vs.\ 726 Bicycle samples) and genuine structural similarity between two vehicles. VLM distinguishes them through semantic reasoning about features such as engine blocks and exhaust pipes that discriminative training does not explicitly encode, achieving 93.2\% recall compared to YOLO's 68.9\%. This reasoning capability extends naturally to novel categories without retraining.


Secondly, YOLO outperforms VLM on pattern-based classes like Crosswalks (90.5\%) and on the negative class Other (66.2\%). VLM, however, exhibits two distinct failures on these categories:
\begin{enumerate}
    \item \textbf{Object-Centric Bias:} VLM demonstrates a strong preference for discrete objects over surface features. It often prioritizes potentially distinct entities (even if secondary) and treats dominant texture patterns like Crosswalks (41.9\% recall vs YOLO's 90.5\%) as contextual background, leading to false negatives.
    \item \textbf{Semantic Over-Interpretation:} VLM struggles significantly with the Other class (21.6\% recall), frequently hallucinating specific labels for generic scenes. Unlike YOLO, which learns a decision boundary for "none of the above," VLM's pre-training forces it to converge to find meaningful concepts in every image. Consequently, it often over-interprets generic roads as Bridges or Crosswalks and background as Palms, unable to accept that an image contains "nothing of interest".
\end{enumerate}

\section{Teacher-Student Distillation}
\label{sec:distillation}

\begin{figure*}[t]
\centering
\subfloat[Taxi: 96\% $\pm$ 8\% recall at N=10.\label{fig:taxi_curve}]{%
    \includegraphics[width=0.32\textwidth]{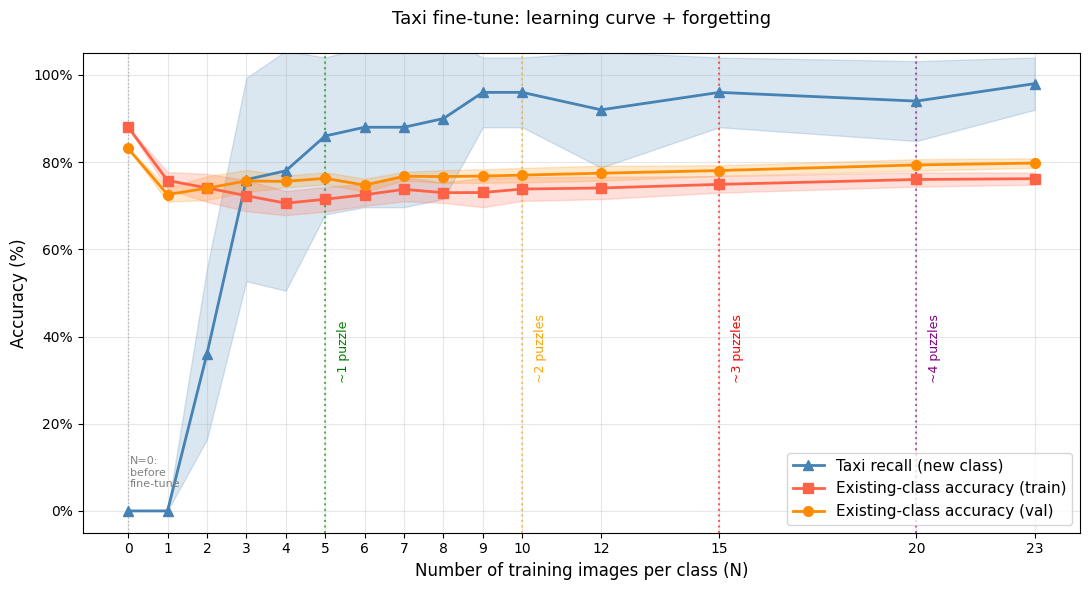}}
\hfill
\subfloat[Tractor: 88\% $\pm$ 10\% recall at N=15.\label{fig:tractor_curve}]{%
    \includegraphics[width=0.32\textwidth]{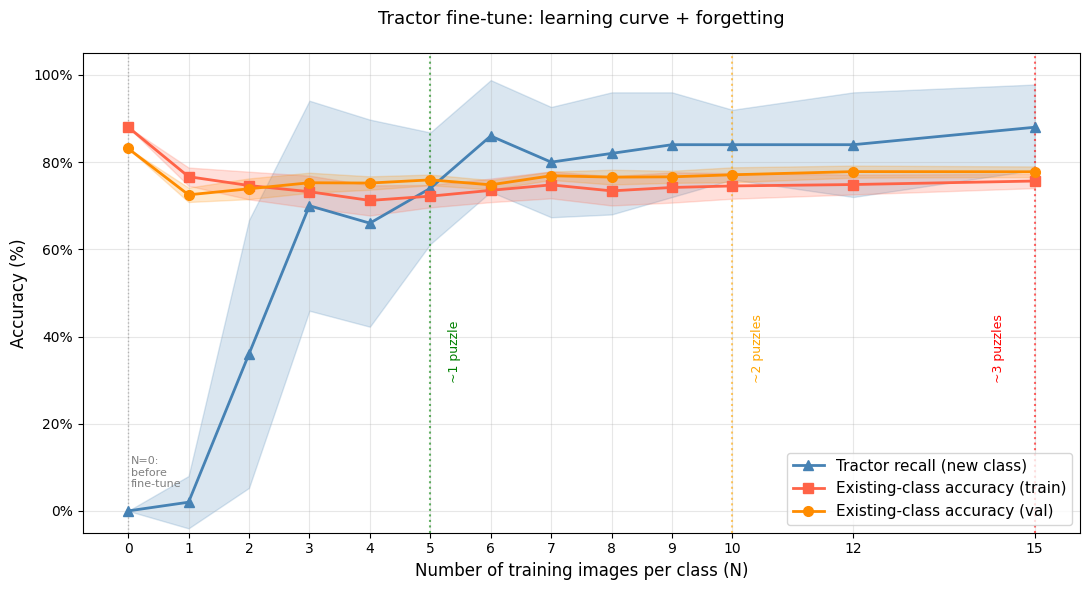}}
\hfill
\subfloat[Boat: 48\% $\pm$ 18\% recall at N=6.\label{fig:boat_curve}]{%
    \includegraphics[width=0.32\textwidth]{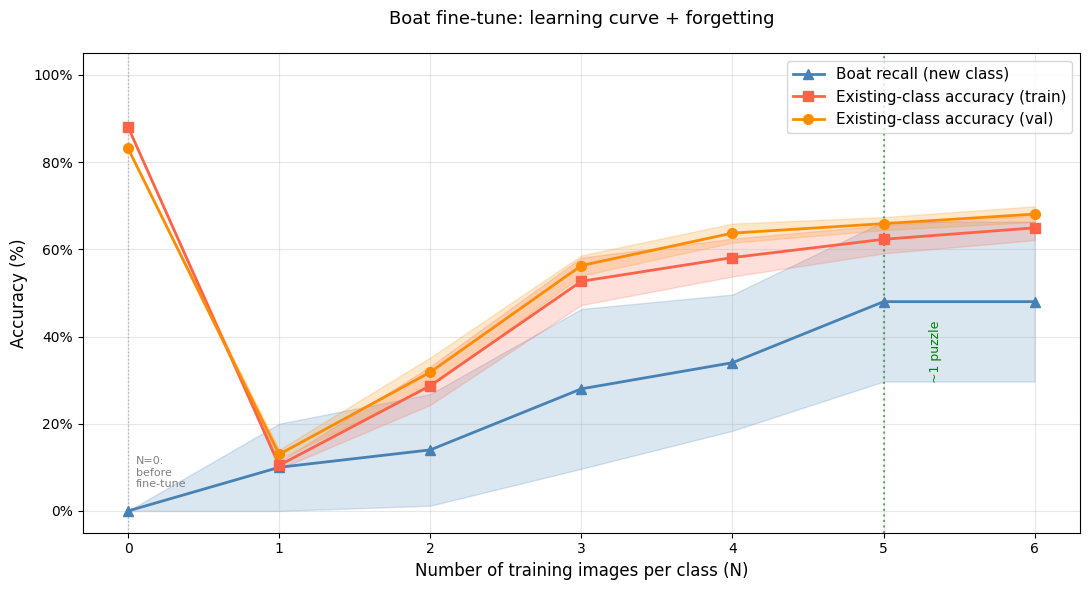}}
\caption{New-class recall as a function of training images ($N$) for each unsupported class added via Experience Replay (10 seeds per point, shaded region shows standard deviation across seeds). Dashed vertical lines indicate approximate puzzle equivalents (e.g., $N$=5 $\approx$ 1 puzzle).}
\label{fig:learning_curves}
\end{figure*}

\begin{table*}[t]
\centering
\caption{Distillation dataset summary and new-class recall (\%, mean $\pm$ std over 10 seeds) after Experience Replay fine-tuning.}
\label{tab:distill_combined}
\small
\begin{tabular}{l ccc cccccc}
\toprule
\multirow{2}{*}{Class} & \multicolumn{3}{c}{Dataset} & \multicolumn{6}{c}{Recall by number of training images (N)} \\
\cmidrule(lr){2-4} \cmidrule(lr){5-10}
& Total & Val & Max Train & N=3 & N=5 & N=7 & N=10 & N=15 & N=23 \\
\midrule
Taxi & 28 & 5 & 23 & 76$\pm$23 & 86$\pm$18 & 88$\pm$18 & 96$\pm$8 & 96$\pm$8 & 98$\pm$6 \\
Tractor & 20 & 5 & 15 & 70$\pm$24 & 74$\pm$13 & 80$\pm$13 & 84$\pm$8 & 88$\pm$10 & --- \\
Boat & 11 & 5 & 6 & 28$\pm$18 & 48$\pm$18 & --- & --- & --- & --- \\
\bottomrule
\end{tabular}
\end{table*}

The hybrid architecture described in the preceding sections demonstrates that VLM achieves high recall on categories where YOLO fails entirely, such as Boat, Taxi, and Tractor. Since VLM already produces correct predictions for these classes, its outputs can serve as labeled training data to fine-tune YOLO, converting expensive VLM reasoning into millisecond-scale reflexes. Prior fine-tuning-based solvers~\cite{plesner2024breaking} depend on a dataset that a human collected and annotated in advance, which fixes their vocabulary to the classes anticipated at training time. Here, the labels come from VLM already present in the pipeline, generated from the very puzzles the solver encounters at deployment, with no human annotation and no pre-collected dataset. The solver therefore extends its own vocabulary to classes it was never deployed with, closing the loop between reasoning and recognition at run time. Knowledge distillation~\cite{hinton2015distillingknowledgeneuralnetwork} transfers knowledge from a capable teacher to a smaller student by training the student to match the teacher's outputs. In our setting, VLM teacher produces hard labels rather than soft probability distributions, reducing distillation to supervised learning:
\begin{equation}
    \mathcal{L}_{\text{distill}} = \sum_{x \in \mathcal{X}} \mathcal{L}_{\text{CE}}(f_S(x), \hat{y}_T(x))
\end{equation}
where $\hat{y}_T(x)$ is VLM's predicted class for input $x$ and $\mathcal{L}_{\text{CE}}$ is the cross-entropy loss. Effective distillation requires a teacher whose predictions are accurate enough to improve the student. For classification, VLM meets this criterion with strong per-class recall that clearly exceeds YOLO on unsupported categories. For segmentation, VLM is too imprecise teacher (Appendix~\ref{app:segmentation}), so we focus distillation and adversarial recovery on classification.

\subsection{Method: Experience Replay Fine-Tuning}

When extending a classifier to support new classes, catastrophic forgetting may occur such that the model forgets previously learned classes when fine-tuned on new data. Experience Replay~\cite{chaudhry2019tiny} (ER) mitigates this by combining exemplars from old classes with new training data. Given an existing model trained on classes $\mathcal{C}_{\text{old}}$ and a new class with $N$ training samples, ER constructs a balanced training set:
\begin{equation}
    \mathcal{D}_{\text{ER}} = \mathcal{D}_{\text{new}} \cup \bigcup_{c \in \mathcal{C}_{\text{old}}} \text{Sample}(\mathcal{D}_c, N)
\end{equation}
where $\text{Sample}(\mathcal{D}_c, N)$ selects $N$ exemplars from each existing class, ensuring equal representation and preventing the model from being dominated by the new class.

We fine-tune the base 13-class YOLO classifier to support each unsupported class (Boat, Taxi, Tractor) as a 14th class using ER to prevent catastrophic forgetting. For each new class with $N$ training images, we include $N$ randomly sampled exemplars per existing class, maintaining balanced representation. We freeze the first 5 backbone layers to preserve learned features, train for 50 epochs with a learning rate of $5\times10^{-4}$, and apply standard data augmentation (color jitter, rotation, scaling, horizontal flip, and random erasing) to mitigate overfitting on small training sets.

For each value of $N$, we run 10 independent trials with different random seeds, each holding out 5 images for validation and sampling different training subsets. With very small datasets, performance can vary significantly depending on which specific images are selected, so running multiple seeds captures this variance. Table~\ref{tab:distill_combined} summarizes the available data per class and presents the mean recall on held-out validation images across 10 seeds.

These experiments use VLM labels exactly as produced, without any human correction or filtering. VLM labeling accuracy on the three new classes is 100\% for Boat and Tractor and 82.1\% for Taxi (5 of 28 images misclassified; see Appendix~\ref{sec:collected_samples}). We deliberately retain the mislabeled Taxi samples rather than filtering them, both because the datasets are already extremely small (11--28 images) and because filtering would require ground-truth knowledge the autonomous pipeline does not have. Consequently, the recall values reported in Table~\ref{tab:distill_combined} already reflect realistic VLM label noise. We quantify the gap between VLM-labeled and ground-truth-labeled training directly in the adversarial recovery experiments (Section~\ref{sec:adversarial}, Table~\ref{tab:self_healing}), where it amounts to a 6--10\,pp penalty.

\subsection{Results}

\paragraph{New-Class Recall.}
To contextualize the $N$ values in terms of real-world CAPTCHA encounters, a single 3$\times$3 classification puzzle typically yields 3--5 positive examples of the target class (not all 9 tiles contain the target object). Thus, $N$=3--5 reflects the data an agent could gather from a single puzzle encounter, while $N$=7--10 corresponds to approximately two puzzles.

Taxi reaches 96\% mean recall by solving 2 puzzles (10 training samples) and exhibits low variance across seeds, indicating that any random subset of taxi images covers enough visual concept. Tractor reaches 88\% at $N$=15 with moderate variance, reflecting moderate intra-class diversity. Boat plateaus at $\sim$48\% mean recall despite using all 6 available training images (approximately one to two puzzle encounters), with high variance depending on which specific images happen to be selected.

The divergence in recall across classes is explained by intra-class visual diversity. Taxi images are almost exclusively yellow sedans on streets. The visual concept is consistent and any random 5 of 28 images look nearly identical. Fine-tuning requires only small, focused weight adjustments to carve out a ``yellow car'' subregion within the existing Car feature space. Tractor images are more diverse in color (green, red, blue) but share a coherent visual structure---large vehicles with proportionally big wheels, typically on farms or roads. The concept is learnable from a moderate number of examples. By contrast, boat images span wildly different visual concepts such as rowboats, speedboats, yachts, sailboats, fishing boats from distinct environments. A yacht shares almost nothing visually with a rowboat. With only 11 total images and a maximum of 6 for training, any random subset fails to represent the full ``Boat'' concept. This finding reveals that intra-class visual diversity relative to sample size is the primary predictor of distillation difficulty in few-shot class extension. For the teacher-student loop to work efficiently, either the new class must be visually consistent or substantially more samples are needed to cover the visual variation.

\paragraph{Catastrophic Forgetting.}
A critical concern in continual learning is whether adding a new class degrades performance on existing classes. Table~\ref{tab:forgetting} reports the existing 13-class accuracy after fine-tuning, compared to the baseline. Note that the baseline here (83.1\%) is higher than the 16-class accuracy in Table~\ref{tab:classification_metrics} (75.48\%) because it excludes four classes (Boat, Stairs, Taxi, Tractor) whose samples were not part of the original YOLO training data~\cite{plesner2024breaking} and are therefore not included in the Experience Replay.

\begin{table}[H]
\centering
\caption{Existing-class accuracy (\%) after fine-tuning at N=5 (mean $\pm$ std, 10 seeds). Baseline: 83.1\% on 13 supported classes.}
\label{tab:forgetting}
\begin{tabular}{lcc}
\toprule
New Class & Accuracy After & Forgetting \\
\midrule
Taxi & 76.3 $\pm$ 1.4 & 6.8\% \\
Tractor & 75.9 $\pm$ 1.2 & 7.2\% \\
Boat & 65.9 $\pm$ 1.5 & 17.2\% \\
\bottomrule
\end{tabular}
\vspace{-2mm}
\end{table}
\noindent Taxi and Tractor fine-tuning produces approximately 7 percentage points of forgetting at $N$=5, a modest cost for gaining a previously unsupported class. Moreover, forgetting decreases as more training samples become available: Taxi forgetting drops from 6.8\,pp at $N$=5 to 3.3\,pp at $N$=23, and Tractor drops from 7.2\,pp to 5.3\,pp at $N$=15. Boat, however, causes 17.2 percentage points of forgetting at $N$=5 and remains high (15.0\,pp) even at its maximum $N$=6. This mirrors the recall results: the model must make broad, destructive weight updates to accommodate Boat's visually dissimilar images under a single class, damaging existing class boundaries in the process.

\section{Adversarial Robustness and Autonomous Recovery}
\label{sec:adversarial}

The preceding sections demonstrate that the hybrid architecture achieves strong classification performance and can expand its capabilities through VLM-guided distillation. A natural countermeasure for the defender is to neutralize this advantage with adversarial perturbations. We investigate whether such attacks succeed and whether the architecture recovers from it.

\subsubsection*{Threat Model} Throughout the paper, the \emph{solver} is our system, which attempts to pass CAPTCHA challenges automatically, and the \emph{defender} is the CAPTCHA operator seeking to block such automation. Note that the roles are inverted relative to the usual convention, since it is the defender who crafts adversarial perturbations and the solver that must withstand and recover from them. We assume the CAPTCHA defender knows that the solver uses a published YOLO classifier~\cite{plesner2024breaking} and can craft adversarial perturbations~\cite{szegedy2013intriguing} against it. The defender is unaware whether a VLM is part of the solver's pipeline, and has no access to the model identity, prompt template, or predictions. In the grey-box escalation experiments (Section~\ref{sec:greybox}), we strengthen the defender further: they know that the solver retrains on adversarial examples and can reproduce its published model, training data, and training procedure. The only elements they cannot reproduce are the order in which the solver shuffles its training batches and the labels its VLM assigns. We investigate whether the solver recovers under iterative attack, and how much its resistance depends on teacher quality versus training independence.

We generate untargeted adversarial examples using Projected Gradient Descent (PGD)~\cite{madry2017towards} with random initialization. Given a classifier $f$ and input $x$ with true label $y$, PGD iteratively finds a perturbation $\delta$ such that $f(x + \delta) \neq y$ with $\|\delta\|_\infty \leq \epsilon$:
\begin{equation}
\small
    \delta^{(k+1)} \!=\! \Pi_{\|\cdot\|_\infty \leq \epsilon} \!\left( \delta^{(k)} \!+\! \alpha \!\cdot\! \text{sign}\!\left(\nabla_{\delta} \mathcal{L}(f(x \!+\! \delta^{(k)}), y)\right) \right)
    \label{eq:pgd}
\end{equation}
where $\alpha$ is the step size and $\Pi$ denotes projection onto the $\ell_\infty$-ball. We use $\epsilon = 4/255$, step size $\alpha = 1/255$, and 10 PGD steps with random initialization in $[-\epsilon, \epsilon]$, crafting individual perturbations per image. We generate adversarial examples for all images across the 13 YOLO-supported classes (excluding Boat, Taxi, and Tractor which YOLO does not support).

The PGD attack achieves 100\% misclassification, reducing YOLO accuracy from 83.1\% to 0.0\% on the validation set. This vulnerability is not specific to Plesner's model --- any YOLO classifier without adversarial training is similarly susceptible, and could recover through adversarial fine-tuning with correct labels. The same adversarial images, however, have minimal effect on VLM. Since the perturbations were optimized against YOLO's gradients, any effect on VLM represents unintentional transfer rather than a targeted attack. On the validation set, VLM accuracy drops from 78.7\% to 75.0\%, a decrease of only 3.7\,pp, consistent with the gradient misalignment between architecturally distinct models~\cite{demontis2019adversarial}. This asymmetry means that even when YOLO is completely defeated, VLM still classifies the adversarial images correctly. The contribution of our architecture is therefore not that it survives PGD while prior work does not, but that this robust VLM serves as a built-in label source for adversarial retraining --- without human annotation or external data collection --- extending the teacher-student paradigm from Section~\ref{sec:distillation} to adversarial robustness.

\subsection{Autonomous Recovery}

To isolate the effect of VLM label noise on recovery quality, we retrain YOLO on balanced adversarial samples labeled either by our cheap open-weight VLM or by ground-truth (GT) labels, an upper bound a strong, expensive model such as GPT-5, Gemini, or Claude would approach. Varying the training budget from a few hundred to $\sim$11,000 samples spread evenly across classes,\footnote{Of the 13 attacked classes, Stairs is excluded because our pipeline pairs each adversarial image with its clean training counterpart, and Plesner~\etal~\cite{plesner2024breaking} did not release the original Stairs training data, leaving 12 classes.} we retrain on both adversarial and clean copies and evaluate on validation adversarial images crafted against the baseline YOLO.

\begin{table}[t]
\centering
\caption{Autonomous recovery: YOLO accuracy after retraining on adversarial samples labeled by VLM or GT. PGD-10, $\epsilon$=4/255.}
\label{tab:self_healing}
\scriptsize
\begin{tabular}{r cc cc}
\toprule
\multirow{2}{*}{Samples} & \multicolumn{2}{c}{YOLO (VLM labels)} & \multicolumn{2}{c}{YOLO (GT labels)} \\
\cmidrule(lr){2-3} \cmidrule(lr){4-5}
& Clean & Adv & Clean & Adv \\
\midrule
0 (baseline) & 83.0\% & 0.0\% & 83.0\% & 0.0\% \\
442 & 72.1\% & 27.8\% & 80.4\% & 32.6\% \\
897 & 74.0\% & 47.8\% & 80.4\% & 57.4\% \\
1,794 & 72.5\% & 57.0\% & 80.1\% & 66.8\% \\
4,498 & 72.7\% & 62.5\% & 79.3\% & 69.7\% \\
10,790 & 73.1\% & 65.0\% & 77.7\% & 71.0\% \\
\bottomrule
\end{tabular}
\end{table}

\begin{figure*}[t]
  \centering
  \subfloat[Single-round recovery: adversarial accuracy vs.\ training samples. VLM labels (red) incur a 6--10\,pp penalty vs.\ GT (green). Robustness plateaus after $\sim$2,000 samples.\label{fig:sample_efficiency}]{%
    \includegraphics[width=0.48\textwidth]{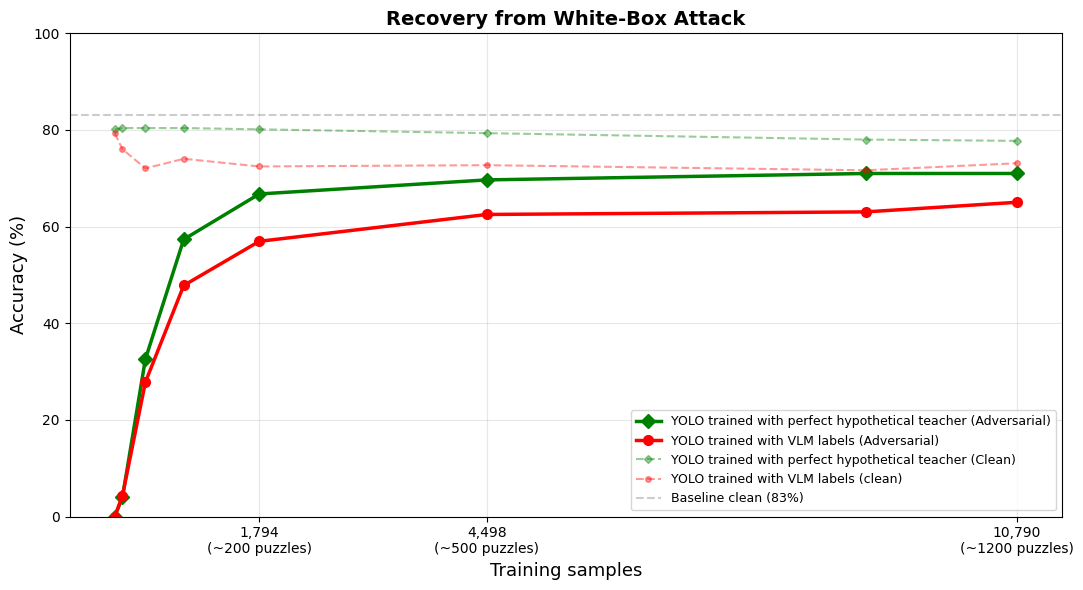}}
  \hfill
  \subfloat[Grey-box escalation over 12 monthly rounds. Pre-recovery accuracy (before each round's retraining) for four solvers, each taught by a perfect oracle or a $\sim$70\%-accurate VLM and trained either \emph{coupled} to the surrogate (same batch order, which the defender can reproduce) or \emph{independent} of it (an order the defender cannot reproduce). The dotted line is the defender's white-box floor. Either an independent batch order or noisy labels decorrelates the solver from the surrogate, and all variants recover after retraining.\label{fig:greybox_escalation}]{%
    \includegraphics[width=0.48\textwidth]{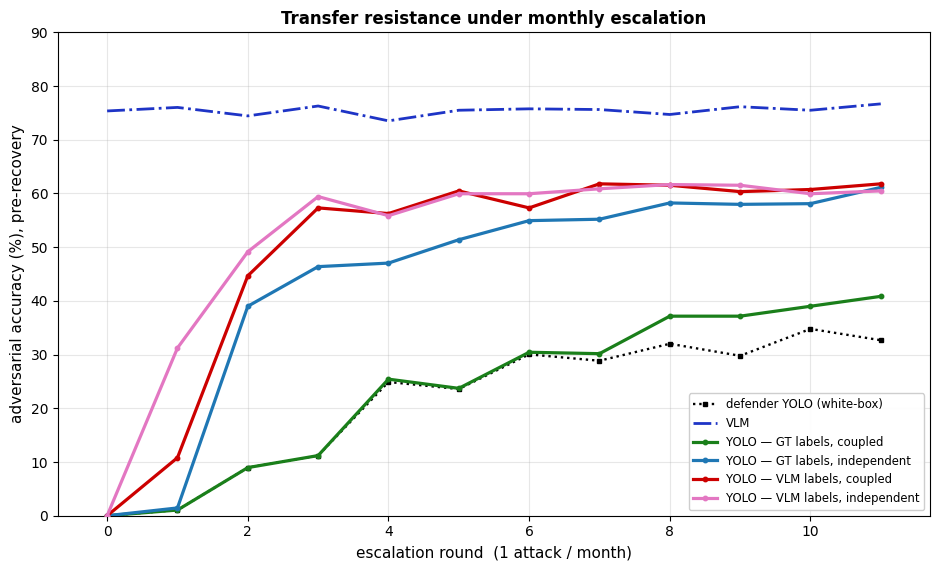}}
  \caption{Adversarial robustness dynamics. (a) Sample efficiency of autonomous recovery under transfer attacks. (b) Iterative grey-box escalation where the defender re-crafts attacks each month.}
  \label{fig:adversarial_dynamics}
\end{figure*}

Table~\ref{tab:self_healing} and Figure~\ref{fig:sample_efficiency} show how adversarial accuracy evolves as the training budget increases.

With GT labels, adversarial accuracy rises from 0\% to 66.8\% using only $\sim$1,800 samples (approximately 200 puzzle encounters). The majority of the gain occurs within the first $\sim$2,000 samples, after which returns diminish, suggesting that a deployed agent could achieve practical robustness from a modest retraining budget.

VLM-labeled retraining incurs a 6--10\,pp penalty in adversarial accuracy compared to GT labels across all training budgets (65.0\% vs 71.0\% at $\sim$11,000 samples), and clean accuracy drops from 83.0\% to $\sim$72--73\% (GT-labeled models retain $\sim$77--80\%). This reflects a standard accuracy-robustness trade-off~\cite{madry2017towards, tsipras2018robustness}, where the retrained model allocates capacity to adversarial patterns at some cost to clean performance.

\subsection{Iterative Grey-Box Escalation}
\label{sec:greybox}

The autonomous recovery experiments above evaluate a single round of retraining against fixed attacks. In practice, the defender re-crafts attacks against an updated surrogate each round, triggering an arms race. We simulate this over 12 rounds, one per month for a year, varying two factors that could shape the solver's resilience, the teacher it learns from and whether the defender can reproduce its training. The two teachers span the cost spectrum. GT labels stand in for an expensive, highly capable model --- the limit a strong proprietary VLM (e.g., GPT-5, Claude) would approach --- while our open-weight 7B VLM, which labels the attacked training images at $\sim$70\% accuracy, is the cheap variant. Every party, defender included, trains with label smoothing (0.1), a standard convention that keeps confidence calibrated (relevant for the cascade under attack). Because the solver builds on the published YOLO classifier~\cite{plesner2024breaking} trained on public data, the defender can replicate its starting point and retraining setup; the one thing they cannot replicate is the order in which the solver shuffles its training batches. Crossing the two teachers with whether the defender can reproduce that batch order gives four solver variants. Each round we record pre-recovery accuracy (before retraining) and post-recovery accuracy (after VLM relabels the attacked images and the solver retrains).

The escalation shows two patterns, one about the solver's recovery after retraining and one about its resistance to the attack before retraining. The first is that recovery is universal (Figure~\ref{fig:greybox_escalation}, Table~\ref{tab:greybox}). Each month the defender's fresh attack degrades the solver, in some months cutting its accuracy on the attacked images to nearly zero before it retrains. Yet once the solver retrains on those images, every variant --- including the coupled perfect-teacher solver, which is nearly identical to the surrogate the defender attacks --- climbs back to 60--64\% accuracy on average (Table~\ref{tab:greybox}). The attack is therefore only a temporary setback. The defender can disrupt the solver each month but cannot prevent it from recovering.

The second pattern concerns how well a solver \emph{resists} an attack before retraining, which depends almost entirely on whether its training trajectory is independent of the surrogate's. Averaged over the year, pre-recovery accuracy is 28.4\% for the coupled perfect-teacher solver, which the defender attacks as a near-clone, but 52.9\% for the same solver trained with an independent batch order --- a 24\,pp gain from trajectory independence alone. VLM-taught solver reaches 58.2\% and is essentially unchanged whether its batch order is coupled (58.2\%) or independent (58.9\%), because its noisy labels already make its trajectory independent of the surrogate, so concealing the batch order provides no additional benefit.

\begin{table}[t]
\centering
\caption{Grey-box escalation, mean pre- and post-recovery accuracy over the 12 monthly rounds. Pre-recovery varies with the solver's independence from the surrogate, while post-recovery is uniform at $\sim$60--64\%.}
\label{tab:greybox}
\small
\setlength{\tabcolsep}{5pt}
\begin{tabular}{ll cc}
\toprule
Teacher & Batch order & Pre & Post \\
\midrule
Perfect (GT) & coupled & 28.4\% & 60.1\% \\
Perfect (GT) & independent & 52.9\% & 64.2\% \\
VLM ($\sim$70\% acc.) & coupled & 58.2\% & 62.3\% \\
VLM ($\sim$70\% acc.) & independent & 58.9\% & 62.3\% \\
\bottomrule
\end{tabular}
\vspace{1mm} \\
{\scriptsize ``Coupled'': the solver shares the surrogate's exact training order, making it a near-clone the defender can attack directly. ``Independent'': the solver shuffles its batches in an order the defender cannot reproduce. Pre and Post are accuracy before and after that round's retraining.}
\end{table}

What resists a transferred attack is a training trajectory that has decorrelated from the surrogate's, and two mechanisms each achieve this: an independent batch order, or VLM's noisy labels. Attack transferability depends on gradient alignment between surrogate and target~\cite{demontis2019adversarial}, and both mechanisms drive that alignment down. Measured against the surrogate at the final round, the coupled perfect-teacher solver stays aligned (0.42), an independent batch order drops it to 0.07, and noisy labels at the \emph{same} batch order drop it further still, to 0.02. The pre-recovery accuracies track this alignment. An independent-batch-order perfect-teacher solver reaches 52.9\% and a same-batch-order VLM solver 58.2\%, both far above the aligned clone's 28.4\%. The two mechanisms are redundant rather than additive --- combining them adds almost nothing (58.9\%), because once a trajectory is decorrelated by either route the surrogate's perturbations no longer transfer. VLM's noisy labels thus genuinely harden the solver, at least as effectively as hiding the batch order would.

This clarifies what the teacher does and does not buy, and the result is counterintuitive. Comparing the two independent-batch-order solvers isolates teacher quality: the perfect teacher yields 52.9\% pre-recovery robustness and the noisy VLM 58.9\%, so the cheap, imperfect teacher not only matches but slightly exceeds the expensive one. Its imperfection is not a liability for robustness but, if anything, an asset --- as the mechanism above shows, noisy labels decorrelate the solver from the defender's surrogate more sharply than clean labels do (0.02 versus 0.07 gradient alignment). What the cheap teacher costs is clean accuracy: VLM-taught solver gives up $\sim$2--6\,pp there while surrendering no adversarial robustness. Teacher quality is thus a clean-accuracy lever, not a robustness lever, and a defender cannot treat the solver's use of a cheap open-weight teacher as a weakness; for hardening, cheap is as good as expensive.

None of this gives the defender a lever. Grey-box knowledge helps only if the solver's training can be reproduced exactly, which VLM labeling prevents at no cost. Deliberately hiding the training order would work equally well, but is a secret that must be set up and maintained, whereas VLM labels come free with solving. Nor do the attacks persist, since retraining restores the solver each round. Throughout, VLM path is itself unaffected, holding $\sim$75\% on the validation attacks because the perturbations target YOLO rather than VLM.

Given that VLM path stays robust, the confidence cascade can serve as a second line of defense under attack, routing the perturbed cells YOLO misreads to VLM --- provided the confidence gate still fires on them. This is where label smoothing earns its place in the training convention. Without it, adversarial retraining drives YOLO's softmax to saturation (28\% of predictions at exactly 1.0), the $\tau=0.70$ gate rarely triggers, and only 4\% of attacked cells escalate. With smoothing, under attack cascade routes 29--46\% of cells to VLM and the cascade lifts pre-recovery accuracy by +11 to +13\,pp --- for VLM-taught solver, from 58.2\% to 70.2\%.



\section{Live reCAPTCHA Evaluation}
\label{sec:live_eval}

To verify how the hybrid solver's performance transfers from static evaluation to live deployment, we evaluated the hybrid solver
against Google's live reCAPTCHA v2 demo page over 100 sessions.

We deploy the system in hybrid mode ($\tau=0.70$)
using the same models, prompts, and FSM as in the static evaluation. The browser was regular Chrome
at 1920$\times$1080 with linear mouse cursor between targets. A session begins at the ``I'm not a robot'' checkbox and
ends in a green tick, representing successful verification. The
FSM does not skip puzzles via pre-attempt reload --- every puzzle
served is attempted, and reloads occur only after verification
failures.

Across 100 sessions comprising 206 distinct puzzles (140 static, 66 dynamic), all 100 sessions passed successfully. 80 sessions cleared without any reload; 18 required one, and 2 required multiple (three and four respectively). At the puzzle level, 181 of 206 puzzles were solved on first attempt (87.9\%); the 25 failures were recovered through reload-and-retry within the same session. The aggregate VLM call rate of 37.1\% is within 7\,pp of the 30\% observed on the static benchmark (Table~\ref{tab:classification_metrics}), confirming that the cascade's routing transfers to deployment. The per-session VLM call rate ranges from 0\% to 75\% (std 15.5\%), reflecting the cascade adapting to per-session difficulty: easy sessions rely entirely on YOLO, hard sessions escalate more to VLM. Dynamic puzzles, which re-render selected tiles after each click, require multiple classification rounds per puzzle as the FSM iteratively re-analyses the grid until no targets remain. This iteration is a deployment-only cost the static evaluation cannot capture, since static images are classified once with no re-analysis. End-to-end, the median time was 17.1\,s per puzzle and 35.2\,s per session.

Plesner~\etal~\cite{plesner2024breaking} report a 100\% session pass rate on the same CAPTCHA family using a Selenium-controlled Firefox browser. However, their setup combines VPN rotation, real-user browser cookies and history, and B\'{e}zier-curve mouse trajectories. Without these helpers, their median required challenges per CAPTCHA degrades from 2 to 5 (cookies removed), 7 (B\'{e}zier replaced with straight lines), and 13 (no mouse simulation), and their no-VPN configuration was blocked entirely after 20 runs. Selenium-controlled browsers expose automation fingerprints that anti-bot systems can detect, which the additional behavioural infrastructure in Plesner's setup compensates for. Their solver also explicitly skips puzzles whose target class their YOLO does not support (Boat, Taxi, Tractor, Stairs). Our screenshot-based approach operates at the operating-system level via mouse and keyboard events, with no DOM access and no browser automation framework, avoiding these fingerprints by design. We achieve a median of 2 distinct puzzles per session and 100\% session pass rate from a private-mode Chrome session on a residential connection, with linear cursor paths, no cookies, and no VPN rotation --- matching Plesner's full-stack median while operating under conditions that block their setup entirely, and handles all 16 target classes via VLM fallback rather than skipping unsupported ones. Halligan~\cite{teoh2025halligan}, the closest comparable VLM-based solver, reports 68\% on reCAPTCHA v2 under a single-attempt-per-challenge protocol. Under that same protocol our first-attempt rate is 87.9\%. On latency, our 17.1\,s median per puzzle is 21\% faster than Halligan's 21.8\,s, which is achieved with self-hosted Qwen-7B-VL rather than proprietary GPT-4o (\$2.40 per 100 puzzles). The cascade naturally limits VLM use to $\sim$37\% of cells, avoiding the per-query API costs of a VLM-only approach.

\section{Conclusion}

This work demonstrates that GUI agents powered by VLMs can serve not only as visual navigators but also as teachers for specialized models. Neither YOLO nor VLM alone is the optimal CAPTCHA solver, but their complementary strengths mean they should collaborate rather than compete. YOLO delivers millisecond-scale inference but fails on unseen categories, while VLM handles any category but at prohibitive cost. By cascading YOLO's confidence-aware predictions with VLM fallback and distilling VLM's knowledge into YOLO over time, our DOM-free hybrid achieves +24.2\,pp macro recall over YOLO alone using VLM for only 30\% of images, while eliminating the need for browser automation frameworks that expose solvers to detection.

The teacher-student paradigm that enables this collaboration extends beyond classification. The same mechanism that teaches YOLO to recognize new object classes from a handful of VLM-labeled puzzle encounters also enables autonomous recovery from adversarial attacks. When PGD perturbations reduce YOLO accuracy to 0\%, VLM relabels the attacked images and the solver retrains, recovering each round over a year-long grey-box arms race in which it is never durably defeated. What determines this robustness is not the teacher's quality. A cheap $\sim$70\%-accurate open-weight VLM hardens the solver at least as well as a perfect oracle, and its noisy labels, far from weakening recovery, decorrelate the solver from the defender's surrogate so that crafted perturbations no longer transfer.

The loop we demonstrate is a general recipe, not a reCAPTCHA v2 tool. Whenever the fast specialist fails --- on an unfamiliar class or an adversarial input --- VLM's reasoning supplies a label, and a handful of such labels is enough to fold that capability into the specialist's reflexes. We show this concretely, extending YOLO to new object classes from as few as one or two puzzle encounters and restoring its adversarial robustness from a couple hundred. Nothing in the recipe depends on v2's particular puzzles, and because a single VLM already reasons across dozens of CAPTCHA families without task-specific training~\cite{teoh2025halligan}, the same distillation applies to any of them --- a stronger teacher would only widen its reach. More generally, whenever an expensive reasoning system operates alongside a fast but limited specialist on deterministic tasks, the reasoner's outputs can progressively bootstrap the specialist's capabilities, converting costly deliberation into hardened reflexes.
\newpage
\bibliographystyle{IEEEtran}
\bibliography{refs}

\newpage
\appendices
\raggedbottom

\section{Open Science}

To support transparency and reproducibility, we provide the following artifacts for evaluation:

\subsection{Available Artifacts}

\begin{itemize}
    \item \textbf{CAPTCHA Solver Application}: Complete implementation with finite-state machine controller, hybrid backend (YOLO/VLM/Hybrid) and pre-trained model weights.

    \item \textbf{UI Detection Model Dataset}: Complete dataset generation scripts for the YOLOv8 UI element detector, with 220 CAPTCHA source images to produce 1,200 annotated training samples with YOLO-format annotations. Due to storage constraints, the generated dataset and background screenshots are not included but can be generated following the provided instructions. Background screenshots can be obtained from the publicly available Kaggle Website Screenshots Dataset~\cite{kaggle_screenshots}. Detailed setup instructions for downloading, preprocessing, and dataset generation are provided in the README of the artifact.

    \item \textbf{Evaluation Dataset}: Due to storage constraints, we include samples only for three object classes (Taxi, Boat, Tractor) that are not available in the dataset by Mandourah~\cite{mandourah2024recaptcha}. Researchers can obtain samples for the remaining classes from the public dataset and run the provided inference scripts to reproduce experimental results. For verification purposes, we include raw prediction JSON files from our original experimental runs.

    \item \textbf{Evaluation Scripts and Notebooks}: Scripts and jupyter notebooks to reproduce experimental results including:
    \begin{itemize}
        \item Classification inference scripts, metrics notebook (\url{classification_metrics.ipynb}), and cascade routing analysis (\url{cascade_routing_analysis.ipynb})
        \item Segmentation evaluation across model sizes
        \item Per-class metric computation
    \end{itemize}

    \item \textbf{Distillation Notebooks} (Section~\ref{sec:distillation}): Experience Replay fine-tuning experiments for each unsupported class:
    \begin{itemize}
        \item Taxi: \url{taxi_finetune_er.ipynb}
        \item Tractor: \url{tractor_finetune_er.ipynb}
        \item Boat: \url{boat_finetune_er.ipynb}
    \end{itemize}

    \item \textbf{Adversarial Robustness Notebooks} (Section~\ref{sec:adversarial}): End-to-end adversarial evaluation pipeline:
    \begin{itemize}
        \item PGD attack generation:\\ \url{01_generate_pgd_attacks.ipynb}
        \item VLM inference on adversarial images:\\ \url{02_vlm_inference_on_adversarial.ipynb}
        \item Sample efficiency analysis:\\ \url{03_sample_efficiency.ipynb}
        \item Grey-box escalation:\\ \url{04_greybox_escalation.ipynb}
    \end{itemize}

    \item \textbf{VLM Backend}: FastAPI server for Qwen-7B-VL model script with deployment instructions, ngrok integration.

\end{itemize}

\subsection{Access Instructions}

The artifact repository, including all source code and experiment notebooks, is available at:

\begin{center}
\url{https://oscilloscope.github.io/captchas-agentic-era/}
\end{center}

\subsection{System Requirements}

\subsubsection{Hardware Requirements}
\begin{itemize}
    \item \textbf{Minimum}: 16GB RAM, 20GB disk space. YOLO-only solver can run on CPU without GPU, though inference will be significantly slower
    \item \textbf{Recommended}: 32GB RAM, 16GB+ GPU VRAM (NVIDIA RTX 3090 or equivalent), 100GB disk space for VLM inference and hybrid mode
\end{itemize}

\subsubsection{Software Dependencies}
\begin{itemize}
    \item \textbf{Operating System}: Any (Windows, macOS, or Linux)
    \item \textbf{Python}: 3.11
    \item \textbf{Browser}: Any (DOM-independent screenshot-based approach)
    \item \textbf{Display Resolution}: The system performs optimally at 1920$\times$1080 resolution. For other display resolutions, browser zoom adjustment may be necessary to ensure reliable UI element localization
\end{itemize}

All Python dependencies with versions are listed in \texttt{requirements.txt}.

\subsection{Installation Instructions}

\begin{enumerate}
    \item \textbf{Create Virtual Environment}:
    \begin{verbatim}
    python -m venv venv
    source venv/bin/activate
    \end{verbatim}

    \item \textbf{Install Dependencies}:
    \begin{verbatim}
    pip install -r requirements.txt
    \end{verbatim}

    \item \textbf{Model Files}: Fine-tuned YOLOv8 weights for UI detection, classification, and segmentation are included in the \texttt{complete\_captcha\_FSM/} directory. For VLM inference, the Qwen-7B-VL model is automatically downloaded from HuggingFace on the first run.

\end{enumerate}

\section{CAPTCHA Solver Implementation}

The \texttt{complete\_captcha\_FSM/} directory contains CAPTCHA solving application that implements the finite-state machine controller and hybrid backend architecture.

\subsection{System Architecture}

The solver implements a human-inspired dual-process approach which combines:
\begin{itemize}
    \item \textbf{YOLO (Fast Reflexes)}: Handles frequent, routine patterns with millisecond-scale inference
    \item \textbf{VLM (Reasoning)}: Intervenes for novel or semantically complex challenges
    \item \textbf{Finite-State Machine}: Coordinates detection, solving, verification, and recovery loops
\end{itemize}

\subsection{Key Components}

\paragraph{Main Entry Point}
\texttt{main.py}: Monitoring loop that continuously checks for CAPTCHA elements on the screen at 2-seconds intervals and invokes the finite state machine when CAPTCHAs are detected between these intervals.

\paragraph{State Machine Controller}
\texttt{captcha\_fsm.py}: Implements the finite-state machine handling CAPTCHA workflow states (detection, checkbox click, puzzle analysis, solving, verification, recovery).

\paragraph{Solver Models}
\url{unified_captcha_processor.py}: Core processor supporting three backends:
\begin{itemize}
    \item \texttt{yolo}: Full YOLO-based detection and classification (default, fastest)
    \item \texttt{llm}: YOLO UI detection with VLM-based classification (requires VLM backend)
    \item \texttt{hybrid}: Confidence-based routing between YOLO and VLM (requires VLM backend)
\end{itemize}

\paragraph{Models}
The solver requires the following pre-trained model files:
\begin{itemize}
    \item \texttt{detection\_model.pt}: YOLOv8 UI element detector (5 classes: captcha area, cell, reload button, submit button, robot checkbox)
    \item \texttt{classification\_model.pt}: YOLO classifier for 3$\times$3 puzzles
    \item \texttt{yolov8x-seg.pt}: YOLOv8x segmentation model for 4$\times$4 puzzles (auto-download on first run)
\end{itemize}

\subsection{Running the Solver}

\paragraph{Basic Usage}
\begin{verbatim}
cd complete_captcha_FSM
python main.py
\end{verbatim}

\paragraph{Command-line Options}
\begin{itemize}
    \item \texttt{--backend \{yolo,llm,hybrid\}}: Choose solver backend.
    \item \texttt{--no-mouse}: Disable mouse movement animations for faster execution. In our experiments, mouse movements did not influence solving success rates
\end{itemize}

\paragraph{Examples}
\begin{verbatim}
# Default (YOLO backend)
python main.py

# VLM backend
python main.py --backend llm

# Hybrid mode
python main.py --backend hybrid
\end{verbatim}

\paragraph{Configuration}
\begin{itemize}
    \item \texttt{config.json}: API URL for VLM backend (required for llm and hybrid modes)
    \item \texttt{config.py}: Mouse movement behavior configuration
\end{itemize}

\subsection{Workflow}

The solver operates in a continuous monitoring loop at 2-seconds intervals:
\begin{enumerate}
    \item \textbf{Detection}: Continuously monitors screen for reCAPTCHA checkboxes using trained YOLOv8
    \item \textbf{Initialization}: Clicks checkbox and waits for CAPTCHA dialog to appear
    \item \textbf{Puzzle Analysis}: Captures CAPTCHA area, identifies grid structure, uses EasyOCR to extract target object and determines puzzle type (classification vs segmentation)
    \item \textbf{Backend Routing}: Selects YOLO or VLM backend based on target object and configured strategy
    \item \textbf{Solving}: Clicks cells identified by the selected model
    \item \textbf{Verification}: Validates success by checking if CAPTCHA disappeared and distinguishes between completion, failure, or puzzle re-appearance
    \item \textbf{Recovery}: Reloads puzzle on failure or handles dynamic puzzles where cells refresh after interaction
\end{enumerate}

\section{YOLOv8 UI Detector Training}

\textbf{Note on Datasets:} Our system uses two distinct datasets for different purposes: (1) the UI Detection Training Dataset described in this section (1,200 samples for training YOLO to localize CAPTCHA UI elements on webpages), and (2) the CAPTCHA Evaluation Dataset described in Section~4 (\classificationN{} classification samples and 204 segmentation puzzles with 16 cells each totaling \datasetN{} evaluation samples across \classesN{} object classes). The UI detector trained here is used to extract CAPTCHA-related regions only.

\subsection{Dataset}

We fine-tuned OmniParser's YOLOv8-based icon detection model~\cite{omniparser} for CAPTCHA-specific UI element detection. The UI detection training dataset comprises 1,200 annotated screenshots across five classes:

\begin{itemize}
    \item \textbf{captcha\_area}: Main CAPTCHA region where images appear
    \item \textbf{cell}: Individual clickable grid cells
    \item \textbf{reload\_button}: Refresh button
    \item \textbf{submit\_button}: Verification button
    \item \textbf{robot\_checkbox}: "I'm not a robot" checkbox
\end{itemize}

The dataset includes 880 CAPTCHA screenshots, 200 robot checkbox samples and 120 negative samples (non-CAPTCHA images) to reduce false positives. We used a 80\%/20\% train/validation split.

\subsection{Dataset Generation Methodology}

To create a diverse training dataset without exposing our model to live CAPTCHA services during development, we synthetically generated training samples using the following procedure:

\begin{enumerate}
    \item \textbf{Background Collection}: We collected diverse webpage screenshots from Kaggle Website Screenshots Dataset~\cite{kaggle_screenshots}, which provides varied visual environments that simulate real-world CAPTCHA placement scenarios across different websites.

    \item \textbf{CAPTCHA Collection}: We collected 220 reCAPTCHA screenshots spanning both 3$\times$3 classification and 4$\times$4 segmentation puzzle types across diverse object categories. These serve as source templates for synthetic generation.

    \item \textbf{Automated Overlay and Annotation}: We programmatically overlaid each of the 220 CAPTCHA images on 4 unique background screenshots at varying positions and scales (0.6$\times$ to 1.5$\times$). The script automatically detects grid type (3$\times$3 or 4$\times$4) and generates YOLO-format annotations for all UI elements including checkbox, captcha area, individual cells and buttons.

    \item \textbf{Manual Verification}: The script also generates verification images with bounding boxes overlaid on the raw screenshots for manual verification.
\end{enumerate}

\paragraph{Dataset Composition} The final dataset comprises 1,200 total samples:
\begin{itemize}
    \item 880 CAPTCHA samples containing classes 0--3 (captcha\_area, cell, reload\_button, submit\_button)
    \item 200 robot checkbox samples containing class 4 (checkbox)
    \item 120 negative samples with empty annotation files to reduce false positives on non-CAPTCHA webpages
\end{itemize}

\subsection{Training Configuration}

Hyperparameters: AdamW optimizer (weight decay 0.0005), learning rate 0.001$\rightarrow$0.01 (cosine annealing), batch size 16, image size 512$\times$512, 150 epochs with early stopping. Data augmentation included HSV adjustments, horizontal flip (p=0.5), scale variation ($\pm$30\%) and mosaic augmentation (p=0.7).

\subsection{Results}

\begin{figure}[H]
  \centering
  \includegraphics[width=\columnwidth]{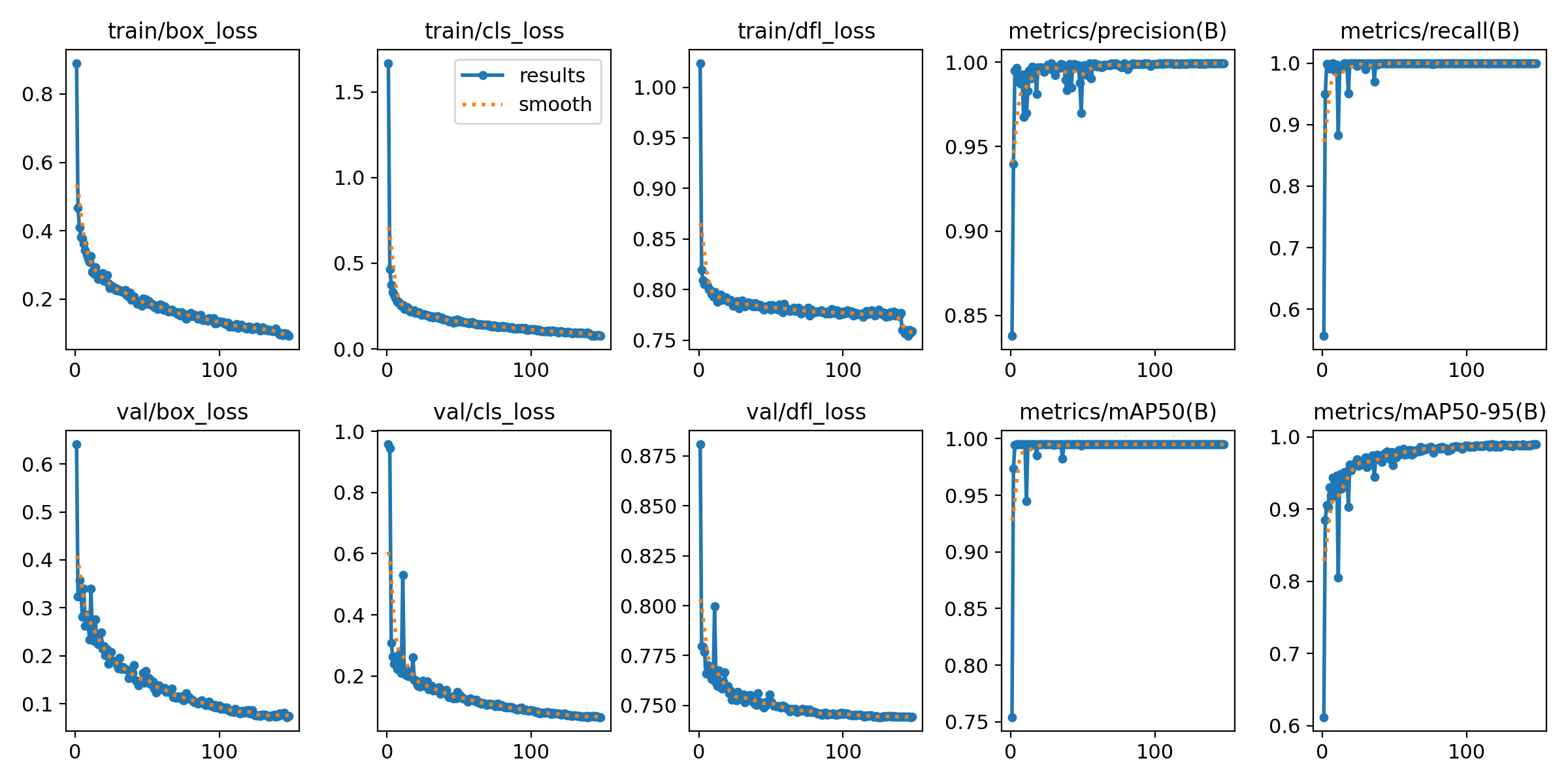}
  \caption{Training and validation metrics over 148 epochs.}
  \label{fig:yolo_training}
\end{figure}

Training converged at epoch 148 (Figure~\ref{fig:yolo_training}). Final validation metrics: Precision 0.999, Recall 1.000, mAP@0.5 0.995, mAP@0.5:0.95 0.990. Inference speed: $\sim$1.1ms per image (GPU).

\section{Metric Generation and Evaluation Methodology}

To ensure reproducibility and transparency, we provide detailed documentation of our metric generation process. All evaluation scripts, notebooks, and raw inference results are organized under the \texttt{experiments/} directory of the artifact repository. The distillation and adversarial experiments are provided as Google Colab notebooks so that all GPU-intensive experiments can be reproduced directly in a browser without requiring local GPU installation or environment configuration.

\subsection{Classification Metrics Generation}

For classification experiments (Section 5.1), we collected model predictions across all \classificationN{} test samples and computed confusion matrices and per-class metrics as the following:

\paragraph{Data Collection}
We ran inference on both YOLO and VLM models to generate prediction files in JSON format. Each prediction file contains:
\begin{itemize}
    \item Image filename and ground truth label
    \item Predicted class label (top-1 prediction)
    \item Model confidence scores (for YOLO only)
    \item Metadata (image dimensions, processing time)
\end{itemize}

\paragraph{Inference Execution}
We ran separate inference scripts for each model configuration:
\begin{itemize}
    \item \url{test_yolo_classification_single.py}: Runs YOLO classifier on individual grid cell images to produce top-1 class predictions
    \item \url{test_llm_classification_single.py}: Runs VLM inference on the same images with numbered class prompts (0--15) to generate top-1 class predictions
\end{itemize}

\paragraph{Metrics Computation}
The JSON prediction files from both models are loaded by the \url{classification_metrics.ipynb} notebook, which filters predictions to the 893-image test set (759 validation images plus 134 collected-class images from Boat, Stairs, Taxi, and Tractor), builds confusion matrices, and computes per-class metrics using standard definitions:
\begin{itemize}
    \item Precision = TP / (TP + FP)
    \item Recall = TP / (TP + FN)
    \item F1-score = 2 $\times$ (Precision $\times$ Recall) / (Precision + Recall)
\end{itemize}

\subsection{Segmentation Metrics Generation}

For segmentation experiments (Section 5.2), we evaluated both YOLO models (in dual-mode: bounding box and mask) and VLM (only bounding box) across 204 puzzles with 3,264 total cell decisions. For the evaluation:

\paragraph{Inference Execution}
We ran separate inference scripts for each model configuration:
\begin{itemize}
    \item \url{test_yolo_dual_mode.py}: Runs YOLOv8n/m/x in both bounding box and mask modes on all segmentation puzzles
    \item \url{test_llm_unified.py}: Runs VLM inference with bounding box output on the same puzzle set
    \item Each puzzle's results are saved in individual directories with ground truth masks and predicted outputs
\end{itemize}

\paragraph{Cell-Level Metrics Calculation}
For each puzzle, we computed cell-level confusion matrix by:
\begin{enumerate}
    \item Loading ground truth binary mask (which cells should be selected)
    \item Loading model predictions (bounding boxes or segmentation masks)
    \item Mapping predicted regions to 4$\times$4 grid cells using 1\% overlap threshold
    \item Computing TP, FP, FN, TN for each cell decision
    \item Aggregating metrics across all puzzles for the same target object class
\end{enumerate}

\paragraph{Per-Class Aggregation}
The per-class metrics script aggregates cell-level metrics by target object class to compute the per-class F1-scores reported in Table~\ref{tab:seg_per_class}. For each class, we apply the following:
\begin{itemize}
    \item Sum TP, FP, FN, TN across all puzzles targeting that object
    \item Calculate precision, recall, and F1-score from aggregated confusion matrix
    \item Compare performance across different model sizes (n/m/x) and configurations (bbox vs mask)
\end{itemize}

\subsection{Segmentation Results}
\label{app:segmentation}

Table~\ref{tab:seg_results} summarizes performance on segmentation puzzles with \segCellsN{} cell decisions from 204 puzzles. YOLOv8x-mask is the strongest single model (86.7\% accuracy; F1=0.814). VLM achieves higher recall (80.1\%) but significantly lower precision (73.9\%) than YOLO masks. As in classification, a hybrid of YOLOv8x-mask + VLM provides the best overall results (88.6\% accuracy; F1=0.850) by combining YOLO's geometric precision with VLM's coverage on rare and unseen categories.

\begin{table}[!htbp]
\centering
\caption{Segmentation performance: YOLO, VLM, and Hybrid.}
\label{tab:seg_results}
\scriptsize
\begin{tabular}{lcccc}
\toprule
Model & Accuracy & Precision & Recall & F1-Score \\
\midrule
YOLOv8n-BBox & 80.2\% & 78.0\% & 66.8\% & 0.720 \\
YOLOv8n-Mask & 82.8\% & 87.0\% & 64.5\% & 0.741 \\
YOLOv8m-BBox & 82.6\% & 75.9\% & 79.8\% & 0.778 \\
YOLOv8m-Mask & 86.4\% & 86.5\% & 76.4\% & 0.811 \\
YOLOv8x-BBox & 83.9\% & 78.1\% & 80.5\% & 0.793 \\
YOLOv8x-Mask & 86.7\% & 87.2\% & 76.3\% & 0.814 \\
VLM (BBox) & 81.6\% & 73.9\% & 80.1\% & 0.769 \\
Hybrid & 88.6\% & 85.6\% & 84.3\% & 0.850 \\
\bottomrule
\end{tabular}
\end{table}

Table~\ref{tab:seg_per_class} presents per-class F1-score performances across YOLO segmentation models and VLM. The results reveal that (1) larger YOLO models outperform smaller ones on supported classes, with YOLOv8x-Mask achieving the highest F1-scores (Bus: 0.901, Motorcycle: 0.898); (2) all YOLO models fail completely (F1=0.000) on non-COCO classes (Crosswalk, Stairs, Taxi) regardless of size; (3) VLM achieves competitive performance with YOLOv8n but is significantly outperformed by larger YOLO models; (4) despite this, VLM provides full coverage across all 8 classes without any CAPTCHA-specific training with F1=0.857 on Taxi, 0.752 on Stairs and 0.685 on Crosswalk---categories where YOLO has zero capability.

\begin{table*}[t]
\centering
\caption{Segmentation Per-Class F1-Scores}
\label{tab:seg_per_class}
\scriptsize
\begin{tabular}{lccccccc}
\toprule
Class & n-BBox & n-Mask & m-BBox & m-Mask & x-BBox & x-Mask & VLM \\
\midrule
Bicycle & 0.721 & 0.752 & 0.715 & 0.745 & 0.683 & 0.707 & 0.710 \\
Bus & 0.867 & 0.888 & 0.880 & 0.905 & 0.894 & 0.901 & 0.825 \\
Crosswalk$^{\dagger}$ & 0.000 & 0.000 & 0.000 & 0.000 & 0.000 & 0.000 & 0.685 \\
Fire Hydrant & 0.667 & 0.679 & 0.857 & 0.925 & 0.892 & 0.906 & 0.756 \\
Motorcycle & 0.742 & 0.775 & 0.813 & 0.868 & 0.843 & 0.898 & 0.792 \\
Stairs$^{\dagger}$ & 0.000 & 0.000 & 0.000 & 0.000 & 0.000 & 0.000 & 0.752 \\
Taxi$^{\dagger}$ & 0.000 & 0.000 & 0.000 & 0.000 & 0.000 & 0.000 & 0.857 \\
Traffic Light & 0.766 & 0.771 & 0.888 & 0.896 & 0.913 & 0.885 & 0.735 \\
\bottomrule
\multicolumn{8}{l}{\scriptsize $^\dagger$Unsupported by YOLO segmentation models} \\
\multicolumn{8}{l}{\scriptsize n=YOLOv8n, m=YOLOv8m, x=YOLOv8x; BBox=bounding box, Mask=segmentation mask.} \\
\end{tabular}
\end{table*}

The macro-averaged metrics favor VLM due to its zero-shot coverage on unsupported classes. However, when aggregating across all 3,264 cell-level decisions (weighted by puzzle frequency), YOLOv8x-Mask achieves superior overall performance: 87.2\% precision, 76.3\% recall and F1=0.814, compared to VLM's 73.9\% precision, 80.1\% recall and F1=0.769. This demonstrates that, for segmentation tasks, while YOLO excels on supported classes which dominate the dataset, VLM provides coverage for rare and unseen categories.

\section{VLM Backend Infrastructure}

To enable VLM inference in our hybrid CAPTCHA solver, we developed a dedicated API backend that serves fine-tuned Qwen-7B-VL GUI Agent. This section documents the backend architecture, deployment options and configuration instructions.

\subsection{Architecture Overview}

VLM backend is implemented as a RESTful API server using FastAPI.

\subsection{Implementation Details}

\paragraph{Model Loading}
The backend uses the HuggingFace Transformers Library to load Qwen-7B-VL model fine-tuned for GUI tasks~\cite{holo2025}. On first run, the model is automatically downloaded from HuggingFace Hub ($\sim$16GB) and cached locally. Subsequent runs load the cached model.

\paragraph{Image Processing Pipeline}
The \texttt{/generate} endpoint accepts:
\begin{itemize}
    \item Image file (PNG, JPG, etc.) as multipart form data
    \item Messages JSON containing the prompt in OpenAI-compatible chat format
\end{itemize}

For each request, the backend loads the uploaded image and records its original dimensions, then resizes it to meet model input requirements. The chat template is applied to format the prompt, followed by tokenization of both text and image inputs. Inference is executed and the generated text response is decoded, then returned to the client.

\paragraph{Response Format}
The API returns a JSON response containing:
\begin{itemize}
    \item \texttt{model\_output}: Raw text generated by VLM
    \item \texttt{debug\_info}: Inference duration in seconds
    \item \texttt{original\_dimensions}: Width and height of uploaded image
    \item \texttt{resized\_dimensions}: Dimensions after smart resize
\end{itemize}

\subsection{Deployment Options}

\paragraph{Local Deployment}
For local testing and development:
\begin{verbatim}
python gui_agent_backed.py
\end{verbatim}
The API runs on \texttt{http://0.0.0.0:8000} and accepts requests from localhost.

\paragraph{Public Deployment via Ngrok}
For remote access, the backend supports ngrok tunneling:
\begin{verbatim}
python gui_agent_backed.py \
  --ngrok-token YOUR_TOKEN
\end{verbatim}
This creates a public HTTPS URL that forwards requests to the local server, enabling VLM inference from remote machines without network configuration. This approach was used during our experiments to separate GPU-intensive VLM runs from the main solver execution.

\subsection{Integration with Main Solver}

The main CAPTCHA solver sends HTTP POST requests to VLM backend's \texttt{/generate} endpoint. For classification tasks, our trained detection model crops individual grid cells and sends them with the classification prompt. For segmentation tasks, the individual cells are concatenated and sent to retrieve bounding box predictions.

All backend code, deployment instructions, and example API calls are included in the \texttt{llm\_backend/} directory of the artifact repository.

\section{Evaluation Examples}

\begin{figure}[H]
  \centering
  \subfloat[]{\includegraphics[width=0.22\textwidth]{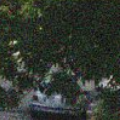}}
  \hfill
  \subfloat[]{\includegraphics[width=0.22\textwidth]{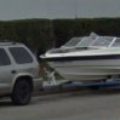}}

  \vskip0.3em

  \subfloat[]{\includegraphics[width=0.22\textwidth]{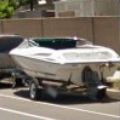}}
  \hfill
  \subfloat[]{\includegraphics[width=0.22\textwidth]{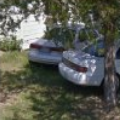}}

  \caption{All four cells are labeled as Car in the ground truth, but our VLM predicts Boat in each case. In (b)--(c), a boat on the road is clearly visible. In (a)--(d), the cars are heavily occluded by environment or appear adjacent to water, making a prediction of Boat a semantically plausible interpretation. These examples illustrate how semantically reasonable predictions can still be counted as errors in our evaluation.}
  \label{fig:boats_inside_car}
\end{figure}

\section{Solver Algorithm}

\begin{algorithm}[H]
\caption{Hybrid CAPTCHA Solver}
\label{alg:pixelcap}
\begin{algorithmic}[1]
\Require screenshot $I$
\State $B \gets \text{detect\_UI\_elements}(I)$
  \Comment{YOLOv8n}
\State $I_{grid} \gets \text{extract\_captcha\_area}(I, B)$
\State $t, type \gets \text{OCR}(I_{grid})$
  \Comment{Target \& type}
\State $R \gets \text{detect\_cells}(I_{grid})$

\If{$type = \texttt{classification}$}
  \ForAll{$r \in R$}
    \If{$t \notin \mathcal{C}_{\text{YOLO}}$}
      \State $y_r \gets \text{VLM\_classify}(r, t)$
    \Else
      \State $y_r, p_r \gets \text{YOLO\_classify}(r, t)$
      \If{$p_r < \tau$} \Comment{$\tau = 0.70$}
        \State $y_r \gets \text{VLM\_classify}(r, t)$
      \EndIf
    \EndIf
  \EndFor
  \State $C \gets \{r : y_r = t\}$
\Else \Comment{Segmentation}
  \State $I_{merged} \gets \text{concatenate}(R)$
  \State $M \gets \text{YOLO\_segment}(I_{merged}, t)$ or
  \Statex \hspace{3em} $\text{VLM\_segment}(I_{merged}, t)$
  \State $C \gets \text{map\_to\_cells}(M, R)$
\EndIf

\State $\text{click}(C)$, $\text{wait}(200\,\text{ms})$

\If{$\text{is\_dynamic}()$}  \Comment{Dynamic puzzle detection}
  \State $I_{new} \gets \text{capture\_screenshot}()$
  \While{$\text{didPuzzleChange}(I_{grid}, I_{new})$}
    \State $I_{grid} \gets \text{extract\_captcha\_area}(I_{new}, B)$
    \State $t, type \gets \text{OCR}(I_{grid})$
    \State $R \gets \text{detect\_cells}(I_{grid})$

    \If{$type = \texttt{classification}$}
      \ForAll{$r \in R$}
        \State $y_r \gets \text{route}(r, t)$ \Comment{Eq.~\ref{eq:cascade}}
      \EndFor
      \State $C \gets \{r : y_r = t\}$
    \EndIf

    \If{$C = \emptyset$}  \Comment{No targets found}
      \State \textbf{break}
    \EndIf
    \State $\text{click}(C)$, $\text{wait}(200\,\text{ms})$
    \State $I_{new} \gets \text{capture\_screenshot}()$
  \EndWhile
\EndIf

\State $\text{click}(\text{verify\_button})$
\State \Return $\text{check\_success}()$
\end{algorithmic}
\end{algorithm}

\section{VLM Prompt Templates}
\label{sec:prompts}

\paragraph{Classification Prompt (3$\times$3 puzzles).}
{\scriptsize
\begin{verbatim}
Choose ONLY ONE number from the following list:
0: Bicycle, 1: Bridge, 2: Bus, 3: Car, 4: Chimney,
5: Crosswalk, 6: Hydrant, 7: Motorcycle, 8: Mountain,
9: Other, 10: Palm, 11: Stairs, 12: Traffic Light,
13: Boat, 14: Taxi, 15: Tractor

Answer with ONLY the number (0-15) of the class
you see with the highest confidence.
Do not include any explanation.
\end{verbatim}
}

\paragraph{Segmentation Prompt (4$\times$4 puzzles).}
{\scriptsize
\begin{verbatim}
Find all {object_name} in this image and create
bounding boxes that encompass the entire {object_name}.

Requirements:
- Include ALL visible parts of the {object_name}
- Capture the complete {object_name} from top to
  bottom and side to side
- Make sure the entire {object_name} is contained
  within the bounding box

For each {object_name}, return coordinates:
{Object_name} 1: [x1, y1, x2, y2]
{Object_name} 2: [x1, y1, x2, y2]

Where (x1, y1) is top-left corner and (x2, y2) is
bottom-right corner.
\end{verbatim}
}

\section{Pipeline Diagrams}

\begin{figure}[H]
  \centering
  \includegraphics[width=0.48\textwidth]{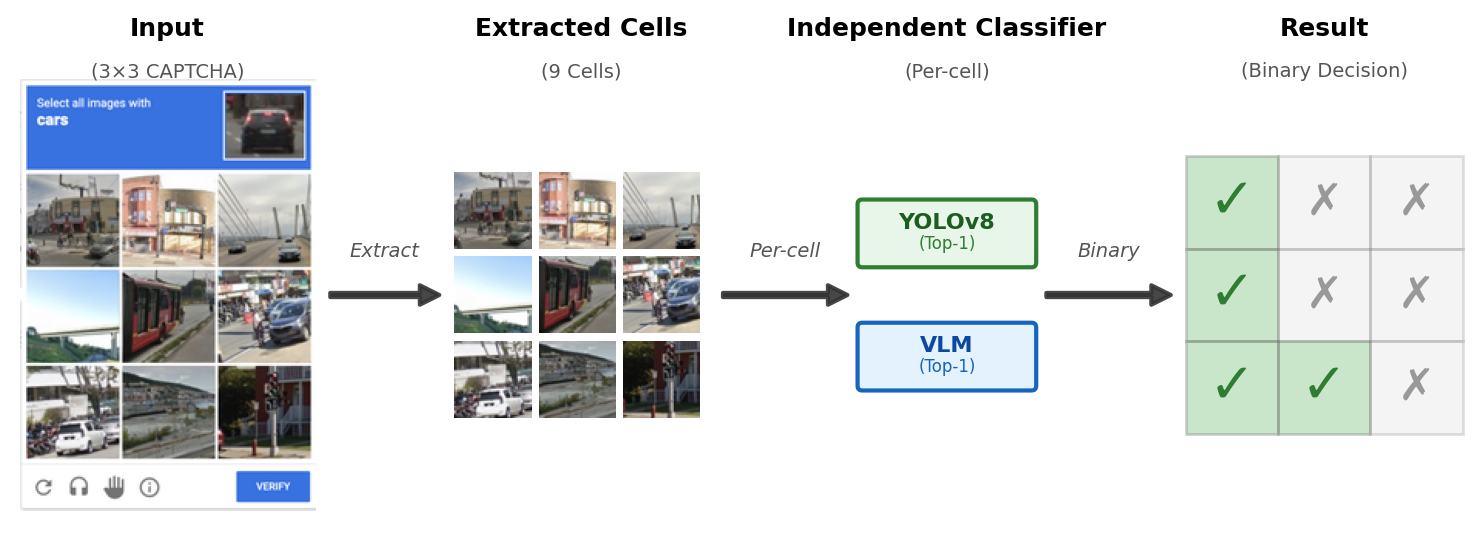}
  \caption{Classification puzzle solving pipeline for 3$\times$3 CAPTCHAs. The system extracts 9 individual cells from the detected CAPTCHA area, classifies each cell independently using either YOLOv8 (top-1) or VLM (top-1) and produces binary decisions indicating which cells contain the target object (cars in this example). Both models operate in top-1 prediction, committing to exactly one class label per cell.}
  \label{fig:classification_pipeline}
\end{figure}

\begin{figure}[H]
  \centering
  \includegraphics[width=0.48\textwidth]{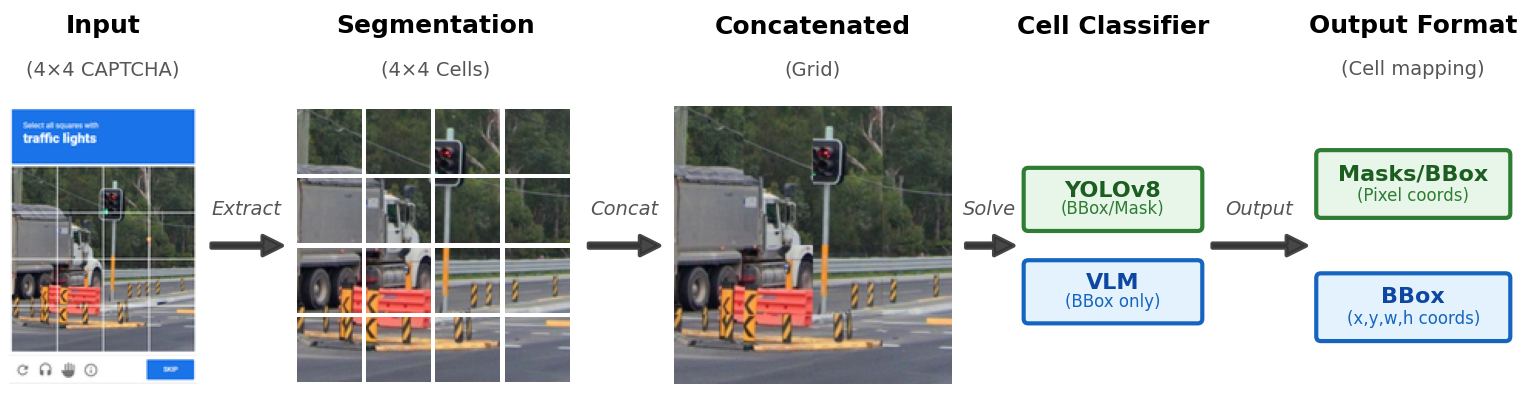}
  \caption{Segmentation puzzle solving pipeline for 4$\times$4 CAPTCHAs. The system extracts and concatenates 16 cells into a unified image, runs either YOLOv8 segmentation (producing pixel-accurate masks or bounding boxes) or VLM (producing bounding boxes only) and maps detected regions back to cell coordinates. Any cell overlapping the predicted region by $\geq$1\% is selected.}
  \label{fig:segmentation_pipeline}
\end{figure}

\section{Collected Class Samples}
\label{sec:collected_samples}
\begin{figure*}[p]
  \centering
  \includegraphics[width=\textwidth]{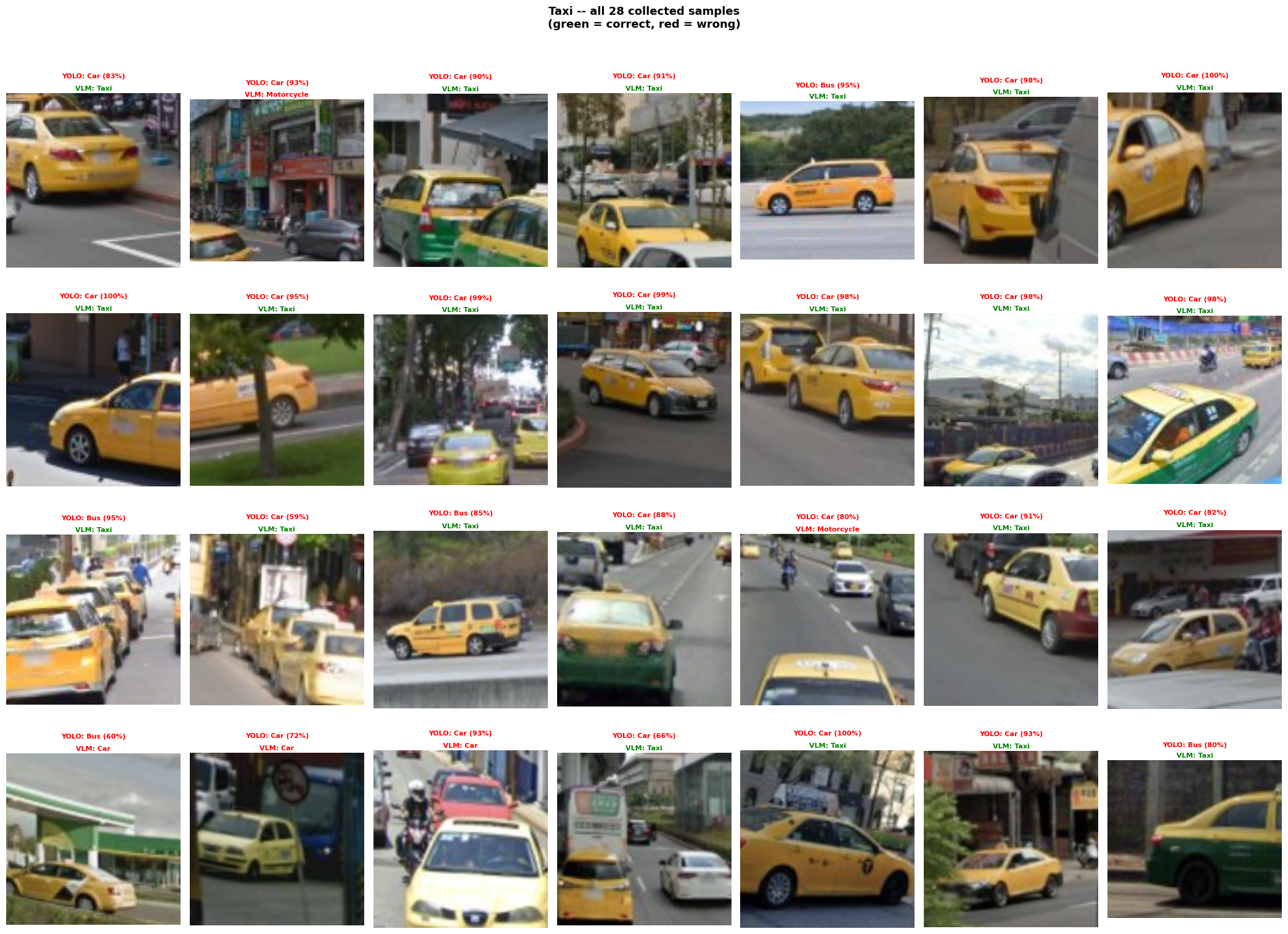}
  \caption{All 28 collected Taxi samples with YOLO and VLM predictions. YOLO misclassifies all images (predominantly as Car), while VLM correctly identifies 23 of these 28 samples (82.1\% recall). The 5 VLM errors are predicted as Car (3) and Motorcycle (2).}
  \label{fig:taxi_samples}
\end{figure*}

\begin{figure*}[p]
  \centering
  \includegraphics[width=\textwidth]{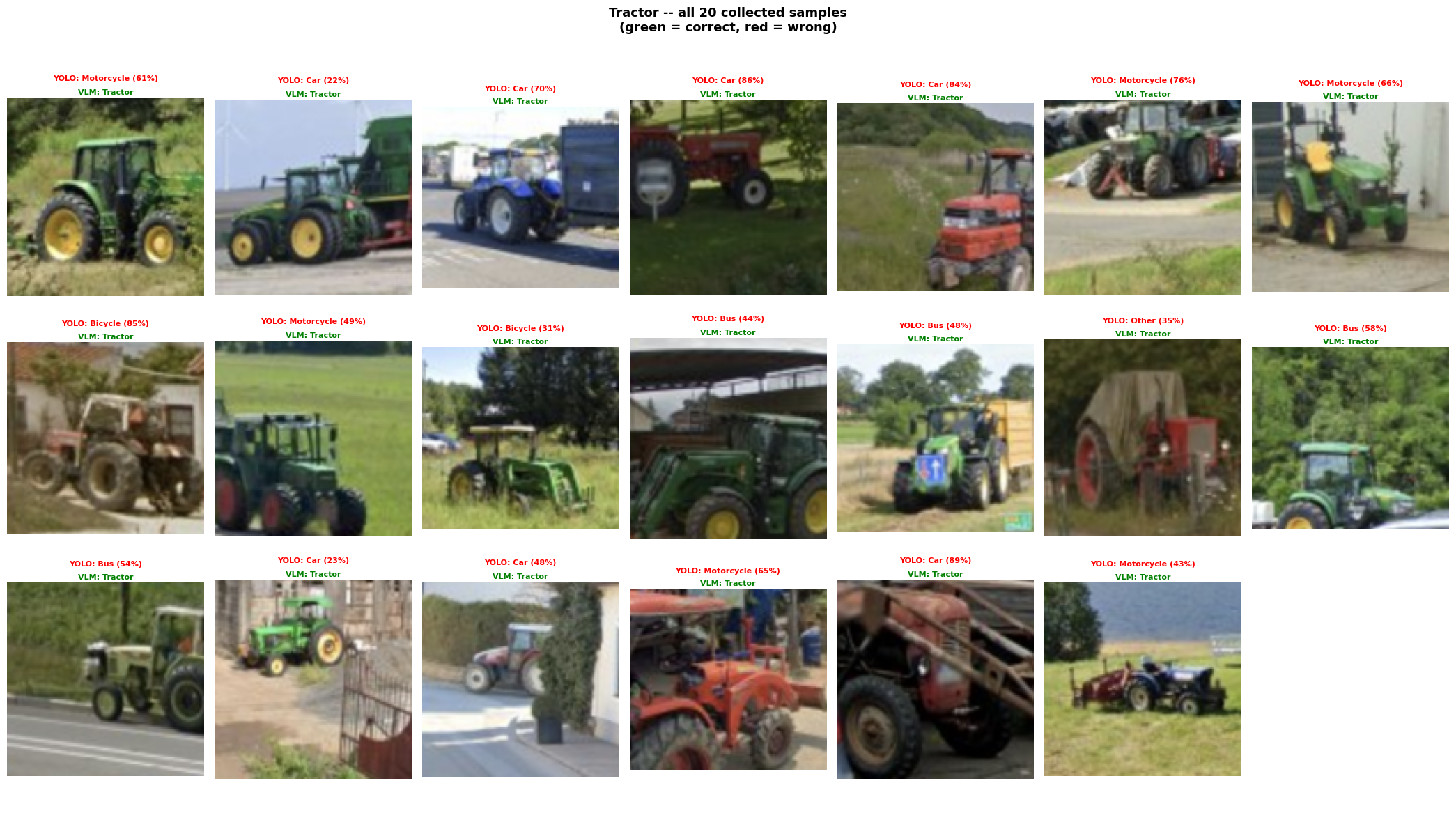}
  \caption{All 20 collected Tractor samples with YOLO and VLM predictions. YOLO misclassifies all images, while VLM achieves 100\% recall.}
  \label{fig:tractor_samples}
\end{figure*}

\begin{figure*}[p]
  \centering
  \includegraphics[width=\textwidth]{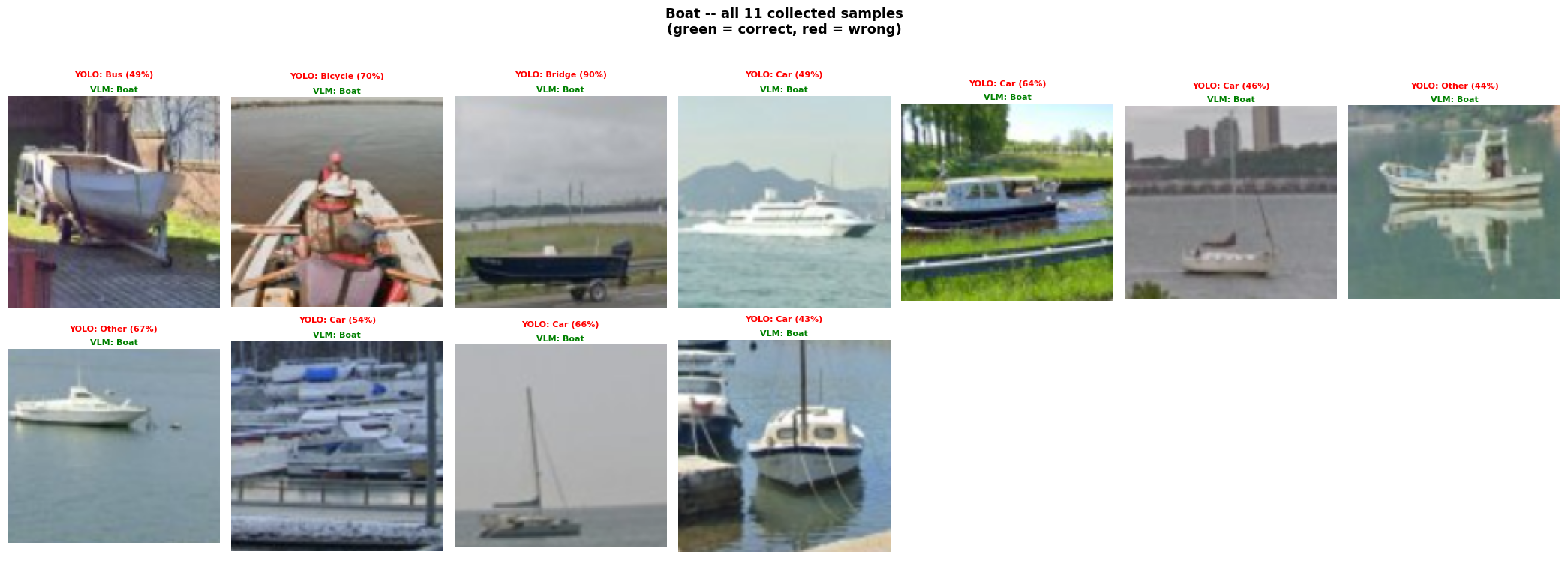}
  \caption{All 11 collected Boat samples with YOLO and VLM predictions. YOLO misclassifies all images, while VLM achieves 100\% recall.}
  \label{fig:boat_samples}
\end{figure*}

Figures~\ref{fig:taxi_samples}--\ref{fig:boat_samples} show all collected samples for the three unsupported classes (Taxi, Tractor, Boat) used in the distillation experiments (Section~\ref{sec:distillation}). Each image is annotated with both YOLO and VLM predictions. Green indicates a correct prediction and red indicates a misclassification. YOLO predictions are always incorrect for these classes since they are outside its supported 13-class. VLM achieves 82.1\% recall on Taxi (23/28), 100\% on Tractor (20/20), and 100\% on Boat (11/11).

\end{document}